\documentclass[iop,apj,twocolumn,twocolappendix,numberedappendix,usenames,dvipsnames]{aastex631}

\usepackage[T1]{fontenc}
\usepackage{amsmath}
\usepackage{amssymb}
\usepackage{ctable} 
\usepackage{graphicx}
\usepackage{natbib}
\usepackage[titletoc,title]{appendix}
\usepackage{ulem}
\usepackage{multirow}
\usepackage{mathtools}
\usepackage{lmodern}
\usepackage{csquotes}
\usepackage{xspace}
\defcitealias{dubey2023}{Paper-I}
\begin{document}

\title{Particles in Relativistic MHD Jets II: Bridging Jet Dynamics with Multi-waveband Non-Thermal Emission Signatures}
\shorttitle{Synthetic Jet Emission}
\shortauthors{Dubey et al.}
\author[0000-0002-8506-9781]{Ravi Pratap Dubey}
\altaffiliation{Fellow of the International Max Planck Research School for Astronomy \& Cosmic Physics at the University of Heidelberg (IMPRS-HD)}
\affiliation{Max Planck Institute for Astronomy, K{\"o}nigstuhl 17, D-69117 Heidelberg, Germany}
\email{dubey@mpia.de,  fendt@mpia.de, bvaidya@iiti.ac.in}
\author[0000-0002-3528-7625]{Christian Fendt}
\affiliation{Max Planck Institute for Astronomy, K{\"o}nigstuhl 17, D-69117 Heidelberg, Germany}
\author[0000-0001-5424-0059]{Bhargav Vaidya}
\affiliation{Indian Institute of Technology Indore, Khandwa Road, Simrol, Indore 453552, India}

\correspondingauthor{Ravi Pratap Dubey}

\begin{abstract}
Relativistic magnetized jets, originating near black holes, are observed to exhibit sub-structured flows.
In this study, we present synthetic synchrotron emission signatures for different lines of sight and frequencies, 
derived from three-dimensional relativistic magneto-hydrodynamic simulations of pc-scale AGN jets. 
These simulations apply different injection nozzles, injecting steady, variable, and precessing jets.
Extending our previous study, here, we have developed a bridge to connect jet dynamics and particle acceleration within 
relativistic shocks with non-thermal radiation dominant in jets.
The emission is derived from Lagrangian particles - injected into the jet and following the fluid - accelerated through diffusive 
shock acceleration and subsequently cooled by emitting energy via synchrotron and inverse-Compton processes.
Overall, the different shocks structures lead to the formation of numerous localized emission patterns - interpreted as jet knots. 
These knot patterns can fade or flare, also as a consequence of merging or Doppler boosting, leading to jet variability.
We find knots with high-enough pattern speed supposed to be visible as superluminal motion $\lesssim 5c$. 
Synchrotron spectra of all jets reveal double-humped structures, reflecting multiple electron populations characterized by the nature 
of underlying shock and their age.
The precessing jet is the most powerful emitter, featuring a spectrum flatter than the steady and the variable jet. 
The emission, although essentially governed by the acceleration through shocks, depends on the cooling history of the particle as well.
Overall, the continuous re-acceleration of electrons through shocks along the jet we found, 
is an essential prerequisite for observing extended jet emission over large time-scales and length-scales.
\end{abstract}


\section{Introduction}
\label{sec:intro}
Jets were first observed, by \citet{curtis1918}, as a thin straight ray of matter connected to a nucleus. 
Since then, relativistic jets have been observed in a variety of astrophysical sources, originating from a deep gravitational 
potential of a compact object \citep{seyfert1943} with strong magnetic field and a disk of accreting matter \citep{hawley2015}.  
These sources include Active Galactic Nuclei (AGNs), X-ray binaries (microquasars), gamma-ray bursts, and young stellar objects. 

Relativistic jets are mostly observed in the radio band through their non-thermal emission \citep[see for example][]{lister2016, jorstad2016}. 
Obviously, the observed features and morphology depend strongly on the the inherent dynamical structure of the jet. 
These structures govern the heating and cooling processes in the jet, and thus determine its radiation losses. 
In addition, the interaction with the environment can play an important role.
Depending on the nature of the source, relativistic effects as well as the inclination towards the line of sight (l.o.s.) to the observer play an essential role. 

As a consequence of the relativistic flow speed, the counter-jet, that is receding from to the observer, is highly de-boosted
in many sources. 
As a result, many jets appear as one-sided core-jet structures with components (so-called jet {\em knots}) that approach the observer with seemingly 
superluminal speed, 
reaching projected velocities of even $\sim 40c$, in particular at parsec scales \citep{lister2016, kim2018, giovannini2018, walker2018m87}. 

Long-term light curves of relativistic jets also show outburst of activities leading to variability of the jet flux on both, long and short 
time scales ranging from minutes to years \citep{schmidt1963, rani2017}. 
In the interesting approach of decomposing the total flux of the jet into several knots, \citet{turler2000} explained this variability as a 
result of ejection of these knots from the jet at various times. 
Every new ejection of  knot then leads to a variability, which is then observed, see e.g. in \citet{lico2022}. 
\citet{clairfontaine2021} studied the interaction of moving shocks and stationary recollimation shocks to explain the flares in radio light curves.
It is therefore interesting to study what leads to the ejection of new knots and how this is connected to the dynamics of the jet. 

The multi-wavelength spectral energy distribution (SED) of the jets, produced by combining the results from multiple wavebands ranging from the radio
to $\gamma-$ray energies, presents a characteristic double-hump structure \citep{tavecchio1998, balokovic2016, prince2021}. 
The lower-energy peak of the double-humped SED, in the radio-infra red regime, is a result of synchrotron emission in the jet. 
The higher energy peak in the X-ray regime, on the other hand, can be explained either through inverse Compton (IC) scattering or synchrotron
emission due to another population of electrons \citep{celotti2001,georganopoulos2006,magic2023}. 

The high-resolution observations also reveal strong, quasi-stationary emission features called {\em knots} e.g. HST-1 in 
M87 \citep{harris2006, nakamura2010} and C7 in BL Lac \citep{cohen2014}. 

A number of numerical studies modelling the synchrotron emission for 3D relativistic hydrodynamic (RHD) jets have been performed 
\citep{gomez1995,aloy2000,westhuizen2019}. 
However, they typically assume a posteriori magnetic field distribution and also adopt a post-processing approach to calculate the emissivity. 
\citet{mimica2009} developed an algorithm to model the real-time transport and evolution of non-thermal particles including radiative losses in RHD jets. 
Radiation transfer modeling of synchrotron emission from special relativistic MHD jets launched from a disk surface was applied by \citet{porth2011}.
A similar approach, but for a 3D general relativistic MHD jets was done by \citet{broderick2010}. 
\citet{fuentes2018} modelled the synchrotron emission for a stationary 3D (special) relativistic magneto-hydrodynamic jet. 

However, these studies do not account for the acceleration of particles, and respectively assume a certain - prescribed - particle energy distribution.
Hence, a fully self-consistent approach, accounting for transport of non-thermal particles, as well as their acceleration and radiative losses is imperative
for a detailed modelling of emission from jets.

In our previous study (see \citet{dubey2023}, Paper-I hereafter) we have primarily focused on the effects of different jet 
dynamics on acceleration of high-energy particles moving along with the jet flow.
We have taken into account the acceleration of electrons as a consequence of shocks using the novel approach invented by \citet{vaidya2018} to model 
diffusive shock acceleration (DSA). 
We have also modeled the cooling of the electron population resulting from radiative losses from adiabatic expansion of the jet, synchrotron radiation,
and IC scattering of background cosmic microwave background (CMB) photons.
Specifically, we had investigated how three dynamically different jet injection nozzles - a steady, a time-variable and a precessing velocity injection -
affect the particle energy distribution.

In the present paper, we model the synchrotron emission and IC scattering of background cosmic microwave background (CMB) photon
from relativistic electrons in these jets to derived synthetic emission signatures, in particular mock radio emission maps for different viewing angles. 
In particular, here we focus on studying the radio observations of the pc-scale jets, and hence on the synthetic {\em synchrotron} emission signatures.

Our paper is structured as follows. 
In Section~\ref{sec:model} we discuss the simulation setup we adopt in the paper,
in particular the dynamical modelling of the jet injection and the modelling of the particle acceleration.
We further discuss the radiative cooling, and also the limitations of our current approach.
In Section~\ref{sec:intensity_maps} we show the synthetic high resolution intensity maps and mock intensity maps. 
The time evolution of the jet intensity, along with pattern speed of knots as well as their superluminal motion is discussed in Section~\ref{sec:int_evolution}. 
We show exemplary light curves in Section~\ref{sec:light_curves}. 
Further, we discuss the spectral energy distribution of the jets in Section~\ref{sec:sed}. 
In Section~\ref{sec:population}, we show the positions of various populations of particles in the jet as well as study their characteristics. 
Finally, in Section~\ref{sec:summary} we summarize our results.

\section{Model Setup and Numerical Specifics}
\label{sec:model}
In the following we briefly discuss the model approach of our present study.
Essentially, we consider the dynamical modeling applying relativistic MHD.
This provides the dynamical variables of the jet flow, which are considered, for each time step, in order to calculate the radiative features.
For a more detailed discussion, we refer the reader to \citetalias{dubey2023} and references therein.

Since the present paper emphasizes on radiative features, we will concentrate on this aspect
in the following.

\subsection{Dynamical Modelling}
\label{sec:dynamical_modelling}
We apply the PLUTO code \citep{mignone2007pluto} to solve the set of (special) relativistic, magnetohydrodynamical (RMHD) fluid equations 
\begin{equation}
    \frac{\partial}{\partial t} \begin{pmatrix}
        D \\ \textit{\textbf{m}}  \\ E_t \\ \textit{\textbf{B}}
    \end{pmatrix} 
    + \nabla \cdot \begin{pmatrix}
        D\textit{\textbf{v}} \\ w_t \gamma_{\rm f}^2\textit{\textbf{vv}} - \textit{\textbf{bb}} +  p_t \bar{{\textbf{{I}}}}   \\ \textit{\textbf{m}} \\ \textit{\textbf{vB}} - \textit{\textbf{Bv}}
    \end{pmatrix}^T = 0
\end{equation}
on a three-dimensional (3D), uniform Cartesian grid. 
Here, $D$ is the laboratory density, 
$\textit{\textbf{m}}$ is the momentum, 
$\textit{\textbf{B}}$ is the magnetic field in the lab frame, 
$E_t$ is the total energy density, 
$\textit{\textbf{v}}$ is the velocity,
$\gamma_{\rm f}$ is the Lorentz factor of the fluid, and 
$\bar{\textbf{I}}$ is the diagonal tensor.

To close the thermo-dynamical equations, we apply Taub-Mathews (TM) equation of state (e.o.s.) where the specific enthalpy $h$ is defined as
\begin{equation}
\label{eq:h_tm}
    h = \frac{5}{2} \Theta + \sqrt{\frac{9}{4}\Theta^2 + 1},
\end{equation}
\citep{mathews1971,Mignone2005} where $\Theta = p/\rho$ is the temperature.

We model magnetized, rotating, one-sided relativistic jets injected from different injection
nozzles (see \citetalias{dubey2023}). 

We employ an initially constant density profile across the domain.
Into this ambient gas a cylindrical \textit{injection nozzle} is placed, 
with a radius $r_{\rm j} =1$, a height $z_{\rm j} =1$, and centered at $x=y=z=0$. 
Note that we adopt a very high - unprecedented - resolution in our study, resolving the jet radius with 25 grid cells. 
This is of key importance in order to capture particle acceleration by numerous shocks that are formed in the domain, as suggested 
by the resolution study we performed in \citetalias{dubey2023}.

The domain outside the injection nozzle initially is defined as the \textit{ambient medium} with density $\rho_{\rm a} = 1000$, 
and being at rest, $\textit{\textbf{v}}_{\rm a}=0$. 
The initial gas pressure $p =0.1$ is constant throughout the domain (including the injection nozzle). 
The magnetic field in the ambient medium is purely vertical, $B_{z,\rm a} = 0.176$, thus $B_{r, \rm a} = B_{\phi, \rm a} = 0$.
Here, $B_{r, \rm a}$, $B_{\phi, \rm a}$ and $B_{z, \rm a}$ are the components of the magnetic field in the ambient medium 
in cylindrical coordinates $r$, $\phi$ and $z$, respectively.
The values of these parameters can be converted from code units (as shown above) to the physical units by using appropriate 
normalisation factors, which we discuss in Section~\ref{sec:simulation_runs}.

We inject an under-dense jet with density $\rho_j =1$ from the injection nozzle.
For the latter we apply the following options, giving rise to 
a (i) time-independent, \textit{steady} jet, 
a (ii) time-dependent, \textit{variable} jet, and 
a (iii) \textit{precessing} jet.
These options are defined by the choice of the velocity profile in the nozzle (see \citetalias{dubey2023}). 
We impose a Lorentz factor along the jet axis $\gamma_c = 10$, and 
a magnetic field along the $z-$axis $B_{zc} = 0.18$. 
With a choice of $\gamma_c = 10$ in our study, we focus primarily on the pc-scale jets.
Resulting from the choice of $B_{zc} = 0.18$ and the magnetic field profile in the jet, we ensure at the jet-ambient
medium boundary in the injection nozzle $B_{z,j} = B_{z, \rm a}$. 
The average value of plasma-$\beta$ parameter in the jet is $0.25$.
With this choice, the value of the other jet parameters are derived applying an equilibrium solution derived by 
\citet{bodo2019} (for more details, see Appendix A of \citetalias{dubey2023}).

\subsection{Particle Injection and Acceleration}
\label{sec:particle_acceleration}
We apply the particle module for PLUTO code \citep{vaidya2018,mukherjee2021} to inject Lagrangian macro-particles from the jet nozzle.
These macro-particles have, at each spatial point, the same velocity as the bulk motion of the fluid.
Thus, the Lorentz factor of the macro-particle is same as the Lorentz factor of the fluid $\gamma_f$ in the underlying grid cell.

Each macro-particle represents an ensemble of non-thermal electrons with an energy distribution following a power-law
\begin{equation}
\label{eq:power-law}
    \mathcal{N}(\gamma) = \mathcal{N}_0 \gamma^{-\alpha}    
\end{equation}
with a chosen {\em initial} power-law index $\alpha = 6$. 
This initial power-law distribution is bound by a lower and a higher cutoff in the Lorentz factor of $\gamma_{\rm{min}}$ 
and $\gamma_{\rm{max}}$, respectively. 
Here, $\mathcal{N}(\gamma)$ represents the number of particles (electrons) per unit volume with a Lorentz factor between $\gamma$ 
and $\gamma + d\gamma$, and $\mathcal{N}_0$ is defined by the number density of electrons $N_{\rm{e}}$  as
\begin{equation}
\label{eq:nmicro}
    \int_{\gamma_{\rm{min}}}^{\gamma_{\rm{max}}} \mathcal{N}_0 \gamma^{-\alpha} d\gamma = N_{\rm{e}}.
\end{equation}
Note that $\gamma$ corresponds to the Lorentz factor of the {\em electrons}, and is different from the Lorentz factor of the {\em fluid} (and of the {\em macro-particle}) denoted by $\gamma_f$.

The electron particle density $N_{\rm{e}}$ is then quantified by {\em choosing} the fraction of equipartition of the energy densities between magnetic field and the radiating electrons
\begin{equation}
    N_{\rm{e}} =  \frac{\epsilon^2}{m_e c^2} \frac{B_{\rm{dyn}}^2}{2} \left(\frac{2-\alpha}{1-\alpha}\right)
    \left( \frac{\gamma_{\rm{max}}^{1-\alpha} - \gamma_{\rm{min}}^ {1-\alpha}}{\gamma_{\rm{max}}^{2-\alpha} - \gamma_{\rm{min}}^ {2-\alpha}} \right)
\end{equation}
where $\epsilon \equiv B_{\rm{eq}}/B_{\rm{dyn}}$,
with $B_{\rm{eq}}$ representing the fiducial magnetic field corresponding to equipartition, 
and $B_{\rm{dyn}} = 6.27$ representing the normalized, average magnetic field in the injection nozzle that is actually applied in our 
simulation (see \citetalias{dubey2023} for a more detailed explanation). 
We choose $\epsilon^2 = 10^{-4}$, such that the initial particle energy is in sub-equipartition with the magnetic field energy.
This implies $\mathcal{N}_0 \sim 1.5 \times 10^{10}$ and  $N_e \sim 0.3$ (in code units) $= 0.003 \ cm^{-3}$.
(we refer to Section~\ref{sec:simulation_runs} for more details on the normalisation that is applied). 

As the jet evolves, these particles fill the computational volume along the jet
as they are advected and are distributed along with the gas flow. 
During their evolution, these particles may encounter shocks and are consequently accelerated as a result of diffusive shock acceleration 
(DSA).

This results in a hardening of electron energy spectrum, depending on the strength of the shock (quantified by the compression ratio), 
and the orientation of the shock (represented by the angle between shock normal and the magnetic field)\footnote{We refer the reader to 
Section 2.4 of \citetalias{dubey2023}, where we discuss the modelling of 
DSA in our simulations following the approach developed in \citet{vaidya2018}}. 
Simultaneously, particles also lose energy as a result of synchrotron and IC-CMB radiation (see next section for details). 
This leads to a continuously changing jet electron energy spectrum with time, determined by the efficiency of acceleration and cooling 
in different sections of the jet. 
\subsection{Radiation: Emissivity \& Intensity Maps}
\label{sec:radiation_modelling}
In addition to being accelerated to higher energies, particles can also lose energy 
due to various physical processes. 
In our study, 
we take into account energy losses due 
to adiabatic expansion of the jet as well 
as radiative losses due to synchrotron emission and IC scattering of background CMB photons.
However, in this paper we are interested in studying the synthetic radio signatures. 
Hence, we focus only on synchrotron radiation and do not show radiation resulting from IC-CMB interaction.

The synchrotron emissivity\footnote{I.e. the radiated power per unit frequency, volume and unit solid angle} in the local co-moving 
frame moving with a velocity {\boldmath $\beta$}
along the l.o.s. $\hat{\textit{\textbf{n}}'}_{\rm los}$, 
emitted by particles (in our case, electrons), distributed isotropically in momentum space, and between a minimum 
and maximum energy $E_i$ and $E_f$, respectively, is
given as 
\begin{equation}
\label{eq:emissivity}
    \mathcal{J}'_{\rm syn}(\nu ', \hat{\textit{\textbf{n}}'}_{\rm los}, \textit{\textbf{B}}') = \frac{\sqrt{3} e^3}{4 \pi m_e c^2} \lvert \textit{\textbf{B}}' \times \hat{\textit{\textbf{n}}}'_{\rm los} \rvert \int_{E_i}^{E_f} \mathcal{N'}(E')F(x)dE',
\end{equation}
\citep{ginzburg1965}, where $\nu$ is the frequency, $\hat{\textit{\textbf{n}}'}_{\rm los}$ is the direction of the line of sight (l.o.s.), \textit{\textbf{B}} is the local magnetic field, $e$ is the charge of electron, $m_e$ is the mass of electron, $c$ is the speed of light, and $E=\gamma m_e c^2$ is the energy of the particle (electron). 
Here, the prime denotes quantities measured in the local, co-moving frame. The function 
\begin{equation}
    F(x) = x \int_x^\infty K_{5/3} (z) dz
\end{equation}
is the modified Bessel function integral where 
\begin{equation}
    x = \frac{4 \pi m_e^3 c^5 \nu '}{3eE'^2 \lvert \textit{\textbf{B}}' \times \hat{\textit{\textbf{n}}}'_{\rm los} \rvert}.
\end{equation}
The emissivity in the co-moving frame given by Equation~\ref{eq:emissivity} is converted
to the emissivity in observer's frame by using the Doppler factor,
\begin{equation}
\label{eq:Doppler_factor}
    \mathcal{D}({\boldsymbol \beta}, \hat{\textit{\textbf{n}}}_{\rm los}) = \frac{1}{\gamma_f (1 - {\boldsymbol \beta} \cdot \hat{\textit{\textbf{n}}}_{\rm los})}
\end{equation}
where $\gamma_f$ is the Lorentz factor of the local fluid element, and hence, of the {\em macro-particle}.
Now, the emissivity in the observer's frame of reference, taking into account relativistic effects, can be given as
\begin{equation}
\label{eq:emmisivity2}
    \mathcal{J}_{\rm syn}(\nu, \hat{\textit{\textbf{n}}}_{\rm los}, \textit{\textbf{B}}) = \mathcal{D}^2 \mathcal{J}'_{\rm syn}(\nu ', \hat{\textit{\textbf{n}}'}_{\rm los}, \textit{\textbf{B}}')
\end{equation}
where
\begin{equation}
    \nu ' = \frac{\nu}{\mathcal{D}} ,
\end{equation}
\begin{equation}
     \hat{\textit{\textbf{n}}}'_{\rm los} = \mathcal{D} \left[ \hat{\textit{\textbf{n}}}_{\rm los} + \left( \frac{\gamma_f^2}{\gamma_f + 1} {\boldsymbol \beta} \cdot \hat{\textit{\textbf{n}}}_{\rm los} - \gamma_f \right) {\boldsymbol \beta} \right], 
\end{equation}
and
\begin{equation}
     \textit{\textbf{B}}' = \frac{1}{\gamma_f} \left[ \textit{\textbf{B}} +  \frac{\gamma_f^2}{\gamma_f + 1} \left({\boldsymbol \beta} \cdot \textit{\textbf{B}} \right) {\boldsymbol \beta} \right]
\end{equation}
\citep{delzanna2006}.
Note, that the emissivity is calculated for {\em each macro-particle} applying Equation~\ref{eq:emmisivity2}, 
and is then interpolated on the underlying Eulerian grid.

This eventually provides the emissivity of cell $i$ as obtained in the observer's frame, $J_{\nu, i}$, which depends on 
the frequency $\nu$, and the inclination to the l.o.s., $\hat{\textit{\textbf{n}}}_{\rm los}$.
Repeating this for all the macro-particles, we get a 3D distribution of emissivity in observer's frame $J_{\nu}$.

In order to produce synthetic two-dimensional intensity maps projected into the plane of the sky (orthogonal to the line
of sight), we need to introduce an observer's frame (denoted by the coordinates $X$, $Y$, $Z$).
In this frame, we define the $Z$-axis along the l.o.s. $\hat{\textit{\textbf{n}}}_{\rm los}$, whereas the $X-$ and $Y-$axes
are in the plane of the sky, and are perpendicular to the line of sight $\hat{\textit{\textbf{n}}}_{\rm los}$. 
Choosing such a frame, we can now integrate the emissivity $J_{\nu}$ along the $Z-$axis, assuming that there in no absorption 
of light as it traverses the medium (hence assuming optical depth $\tau = 0$), to get specific intensity (or surface brightness) as 
\begin{equation}
\label{eq:intensity}
    I_{\nu}(X, Y) = \int J_{\nu}(X, Y, Z) dZ
\end{equation}
For simplicity, when plotting emission maps at certain evolutionary time, we exploit a fast-light approximation in the above integration.
Hence, we do not account for the light travel time between two locations of the emission source along the l.o.s. 
Finally, we obtain the net flux at frequency $\nu$ as
\begin{equation}
\label{eq:flux}
    F_{\nu} = \frac{L_{\nu}}{4 \pi D^2} = \frac{1}{D^2}\int \int I_{\nu}(X, Y) dX dY
\end{equation}
where $L_{\nu}$ is the specific luminosity ({\em i.e.} radiated power) at a particular frequency $\nu$ and $D$ is the distance to the astrophysical source of interest. 
The integration in Equation~\ref{eq:intensity} and \ref{eq:flux} will be performed over the respective region of interest.

\subsection{Limitations of the approach}
\label{sec:limitations}
While we think that our approach of combining relativistic MHD simulations with particle acceleration and energy losses by 
radiation in different frequency bands, is yet unique and un-precedented, we nevertheless want to mention a number of limitations involved.
These limitations seem minor at this stage, but definitely deserve future consideration.
In the following we briefly specify a few critical points.

(i) In the MHD simulation, shocks are resolved by only 3 grid cells. 
This is typical for Godunov-type numerical scheme with linear reconstruction, as it is only second-order accurate in space.

(ii) The orientation of each shock front is actually calculated. However, for the calculation of post-shock power-law 
index of the electron energy distribution, we only consider the asymptotic limits of either a parallel or a perpendicular
relativistic shock, based on whether the angle between the magnetic field and shock normal is either smaller or greater than $45^{\circ}$, respectively. 
This is necessary for us because the analytical expression for post-shock particle power-law indices are only available for these limits.

(iii) If a shock is categorised as quasi-parallel, the post-shock power-law index is calculated by analytical estimates from 
\citet{keshet2005}, assuming an isotropic distribution of electrons.
If a shock is categorised as quasi-perpendicular, the post-shock power-law index is calculated by the analytical estimates from \citet{takamoto2015}.

(iv) In case of quasi-perpendicular relativistic shocks, the small angle scattering has been found to be the dominant mechanism 
for accelerating electrons. 
Then, the ratio $\mathcal{G}$ of gyro-frequency and scattering frequency is proportional to energy $E$ of particle \citep{kirk2010, sironi2013}.
In our prescription, however, we take $\mathcal{G}$ to be constant and $= \sqrt{2}$. 
With this choice, we categorize all the shocks as either parallel or perpendicular as described in point (ii) above.

(v) Due to complex nature of multiple relativistic shocks, 
compression ratios can be achieved that are larger than the theoretical limit for RMHD shocks.
In such a case, we set the spectral index of the post-shock particle energy spectrum to the maximum possible limit of 2.23 \citep{kirk2000}.

(vi) We (safely) assume that the electron acceleration time scale is very small compared to the dynamical time scales in the simulation. 
As a result, the electrons in a macro-particle are accelerated instantly.

(vii) We finally assume an optically thin configuration when we calculate the intensity, thus a medium between the source and observer that is fully transparent. 

(viii) We adopt a test particle approach with no feedback from the particles to the fluid or shock structure.
The energy gained by electrons after shock acceleration is not {\em taken out} from the fluid.  
As a result non-linear effects of diffusive shock acceleration \citep{cristofari2022} are not taken considered here. 

(ix) We only consider DSA as accelerating mechanism for the electrons. 
However, acceleration of electrons may result from other processes as well, such as Fermi second order (stochastic) acceleration, which we do not consider here. 
Studies comparing the impact of DSA and Fermi second order particle acceleration have shown that these mechanisms can complement each other resulting in diffuse X-ray emission along localized bright spots  \citep[see e.g.,][]{Kundu2021, kundu2022, wang2023}. 

(x) We consider only synchrotron radiation, IC scattering of background CMB photons, and the adiabatic expansion of the jet
as cooling mechanisms in the electron transport equation. 
Here, we do not account for the synchrotron self-absorption of low-frequency emission in our emission spectra. 
Considering the same in our modelling would suppress the low-frequency flux below a cutoff frequency which depends on the magnetic 
field, electron density and the geometry of the source. 
For example, 3C\,84 is opaque below 20 GHz, where as the flux from 3C 48 is suppressed only below 100 MHz \citep{kellermann1988}.

(xi) The macro-particles in our approach are Lagrangian in nature. 
Hence the Lorentz factor $\gamma$ of the macro-particle is same as that of fluid $\gamma_f$. 
While this assumption may not change our result significantly in the case of AGN jets, in internal shock models \citep{piran1999} 
the relative Lorentz factor between internal shock shells may be important for particle acceleration.

\subsection{Simulation Runs}
\label{sec:simulation_runs}
\begin{deluxetable}{cc}

\tabletypesize{\small}
\tablecaption{Normalization Units \label{tbl:normalization_units}}
\tablehead{
\colhead{Parameter} & \colhead{Conversion Factors}
}
\startdata
$l_0$     &    $0.5 \ {\rm pc}$ \\
$v_0$  &  $2.998 \times 10^{10} \ \rm{cm\,s}^{-1}$ \\
$\rho_0$  & $1.66 \times 10^{-26} \ \rm{g\,cm}^{-3}$ \\ 
\noalign{\smallskip}\hline \noalign{\smallskip}
$N_0$ & $0.01 \ {\rm cm^{-3}}$ \\
$t_0$  & $1.63 \ \rm{yr}$ \\ 
$B_0$  & $13.69 \ \rm{mG}$  \\ 
$p_0$  & $1.49 \times 10^{-5} \ \rm{dyne\,cm}^{-3}$ \\
$T_0$  & $5.41 \times 10^{12} \ \rm{K}$  \\
\enddata    
\tablecomments{Conversion factors from code units to physical scales.
Shown for the three basic parameters: length $l_0$, speed $v_0$ and density $\rho_0$, along with the derived normalization 
factors for particle density $N_0 = \rho_0 / m_{\rm u}$, time $t_0 = l_0/v_0$, magnetic field $B_0 =  v_0 \sqrt{4 \pi \rho_0}$,
pressure $p_0 = \rho_0 v_0^2$, and temperature $T_0 = \mu m_{\rm u} v_0^2/2 k_{\rm B}$. 
Here, $m_{\rm u}$, $\mu$ and $k_{\rm B}$ denote the atomic mass unit, the mean molecular weight and the Boltzmann constant, respectively.
}
\end{deluxetable}
Since the conservative RMHD fluid equations are scale-free, results from corresponding simulations can  be converted, in principle, 
to any astrophysical scale using the appropriate normalisation factors. 
However, adding source and sink terms (for example DSA and radiative losses) requires a proper dimensional treatment.
Hence, modelling radiation from such simulation necessitates a scaling of parameters in physical units.
Consequently, we have used the same normalisation factors we introduced in Table 1 of \citetalias{dubey2023}. 
We reproduce the same here in Table~\ref{tbl:normalization_units} for convenience of the reader\footnote{For details on normalisation,
we refer to Section 2.2 of \citetalias{dubey2023}.}.
With this normalization, the time in the code units can be defined as $ t= \Tilde{t}/t_0$, where $\Tilde{t}$ is the time in physical units. 
The same applies to all other variables.

Thus, while the kinematics of RMHD simulations which we here apply to pc-scale jets could be applied also for {\em e.g.} kpc-scale jets, 
physical variables such as density or magnetic field would be very different, and thus the jet radiation field. 
In addition, certain radiative loss processes that may be important at pc-scales may not be dominant at kpc scales and vice-versa.

We have performed multiple simulation runs in order to study the effect of various 
physical parameters on the emission signatures of the jet. 
We have chosen to investigate the impact of the following dynamical parameters:
(i) various inclinations of the l.o.s. $\hat{\textit{\textbf{n}}}_{\rm los}$, implying different viewing angles of the jet, and 
(ii) different jet injection mechanisms (steady, variable and precessing injections). 
Additionally, we also investigate the effects of the jet magnetic field strength, parameterized by the (initial) magnetic field along the $z-$axis (i.e. the jet axis) $B_c$.

We identify and differentiate simulations with different parameters through their unique identifiers.
The letters in the identifier refer to the nature of injection {\em viz.} {\em Std}, {\em Prc}, and {\em Var} for steady, precessing, and variable injection, respectively. 
The digits refer to the inclination from the l.o.s. $\hat{\textit{\textbf{n}}}_{\rm los}$ in degrees. 
Thus, for example, identifier {\em Prc45} refers to a precessing jet with inclination from the l.o.s. $\hat{\textit{\textbf{n}}}_{\rm los} = 45^{\circ}$.

We note that these simulation runs represent variations of the simulations we performed in \citetalias{dubey2023}, 
which we here identify as {\em Std90}, {\em Prc90}, and {\em Var90} for the steady, precessing and variable injection, respectively. 
The steady jet nozzle  injects jet material {\em continuously} along the jet axis, 
with $v_{z,Std} \simeq 0.995c$ corresponding to a Lorentz factor $\gamma_{\rm f} = 10$. 
In addition, $v_{r} = 0$, and $v_{\phi}$ is calculated applying the equilibrium profiles following \citet{bodo2019} 
(see Appendix A in \citetalias{dubey2023} for details).

The variable jet, on the other hand, follows a {\em time-dependent injection} with 
$v_{z,Var}(t) = v_{flr} + v_0 \cos^2(\omega t)$, where $v_{flr}=0.8 v_{z,Std}$ is a (constant) floor velocity, 
$v_0$ is the amplitude of the variable component of $v_{z,Var}$, and $\omega$ is the frequency of the variation. 

The injection nozzle of the precessing jet is more complex with {\em all three velocity components changing with time}. 
The velocity vector of the precessing jet varies as $\textbf{v}_{Prc} = \mathbb{T}_z \mathbb{T}_x \textbf{v}_{Std}$, 
where
\begin{equation*}
    \mathbb{T}_x = \begin{pmatrix}
              1 & 0  & 0 \\ 
              0 & \cos{\psi}  & -\sin{\psi} \\ 
              0 & \sin{\psi}  & \cos{\psi}
           \end{pmatrix},\,\,\,
        \mathbb{T}_z = \begin{pmatrix}
               \cos{\omega t} & -\sin{\omega t}  & 0 \\ 
               \sin{\omega t} & \cos{\omega t}  & 0 \\ 
               0 & 0 & 1
           \end{pmatrix}.    
\end{equation*}
Here, $\psi = 10^{\circ}$ is the opening angle of the precession cone and $\omega$ is the angular frequency of the precession. 
For both, the variable jet as well as the precessing jet, 
we choose the same angular frequency $\omega$ such that the period $\mathcal{P} = 2 \pi / \omega = 5$.

Additionally, in order to study the nature of emission at different frequencies in the radio band and to produce the spectral 
energy distribution (SED, also called emission spectrum) of the jet, we have performed each simulation run at twenty one different
frequencies ranging (distributed nearly equidistant in $\log$ space) from $0.1$\,MHz up to $10^{14}$\,GHz (413 MeV).

\begin{figure}
    \includegraphics[width=0.9\linewidth]{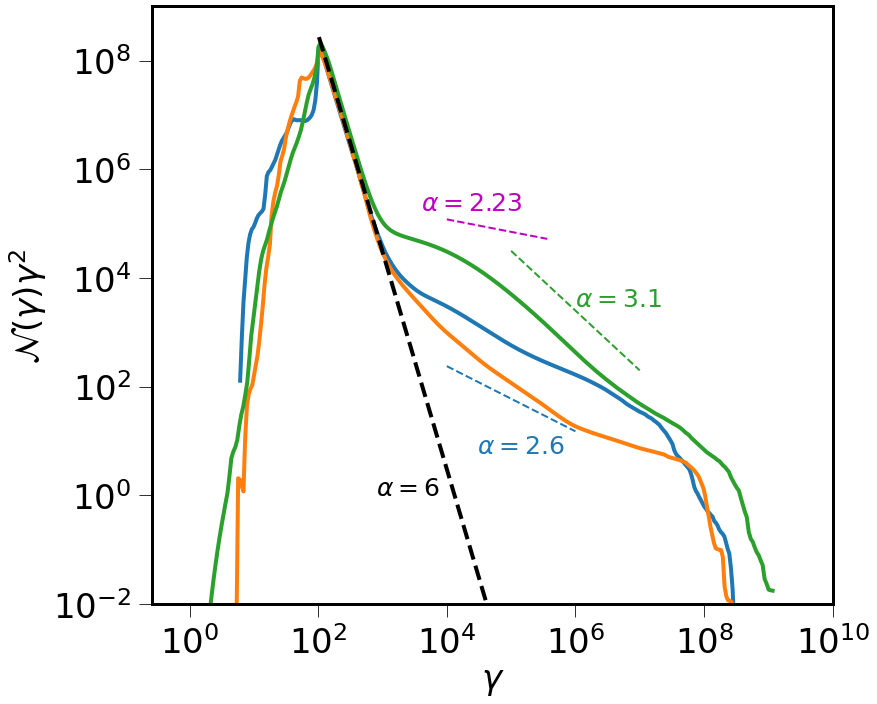}
    \caption{Electron energy spectrum of all particles in the domain for the simulation run \textit{Std90} (in blue), \textit{Var90} (in orange) and \textit{Prc90} (in green) at $t =50$. 
    The injected spectrum for each particle normalized to the total number of particles at $t=50$ is shown by the black dashed line. 
    The blue and green dashed lines show the slope of spectra with power-law index $\alpha = 2.6$ and $3.1$, respectively, for comparison with the particle spectra. 
    The magenta dashed line shows the asymptotic limit of $\alpha = 2.23$ for ultra-relativistic shocks.
    Taken from \citepalias{dubey2023}.
    }
    \label{fig:par_spec}
\end{figure}

For convenience of the reader, we reproduce in Figure~\ref{fig:par_spec} the electron energy spectra of the steady jet {\em Std90}, 
the variable jet {\em Var90}, and the precessing jet {\em Prc90} from \citepalias{dubey2023} at time $t=50$. 
We note here that the maximum achievable Lorentz factor $\gamma$ of the electrons in all the three jets in our study is close to $10^8$. 
This limit is determined by the values of input parameters {\em e.g.} the magnetic field, the bulk Lorentz factor $\gamma_f$ of the 
jet etc., which are same for all the jets we investigate.
We refer the reader to \citepalias{dubey2023} for a detailed discussion of the electron energy spectra of different jet sections.

\section{Intensity Maps}
\label{sec:intensity_maps}
In this section we present and discuss the 2D maps of the specific intensity $I_{\nu}$ as projected onto the plane of the sky for the different jet nozzles, and also for different viewing angles.
The intensity is calculated by integrating the local 3D emissivity of each grid cell along the line of sight (here defined as $Z-$axis) as described in Equation~\ref{eq:intensity}. 
With this, the $X-Y$ plane represents the plane of the sky\footnote{Note that the $X-Y-Z$ coordinate system is different from the $x-y-z$ coordinates applied for the MHD simulations, with latter representing a fixed frame of reference irrespective of the line of sight.}. 

We scale the $X-Y$ coordinate system such that the length of one pixel represents $0.02$\,pc. 
Considering a far away jet source at a distance $D$ (in pc), this unit length corresponds to $(0.02 \times 180 \times 3600)/(\pi D)$\,arcsec.
For M87 with ($D = 16$\,Mpc) our simulation grid resolution corresponds to a resolution on sky of $\simeq 0.258$\,mas. 

In Section~\ref{sec:hr_maps} we present the high-resolution {\em pure} intensity maps from our simulations (as limited by the grid resolution) whereas in Section~\ref{sec:blurred_maps} we show the {\em mock} maps after convolving the pure maps with 
a Gaussian beam.
\begin{figure*}
    \includegraphics[width=0.9\linewidth]{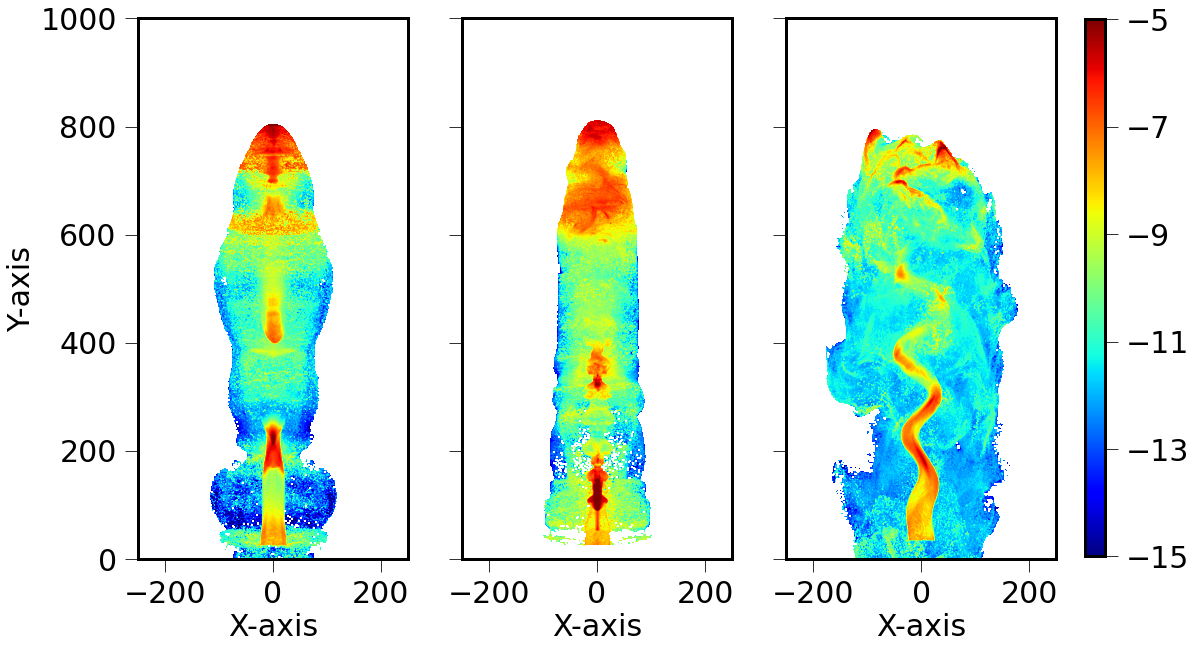}
    \includegraphics[width=0.9\linewidth]{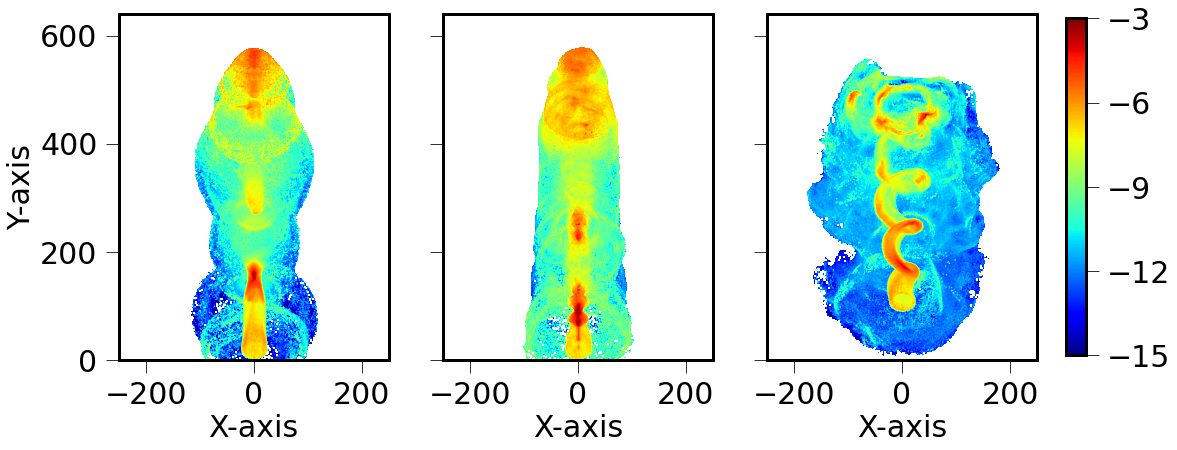}
    \includegraphics[width=0.9\linewidth]{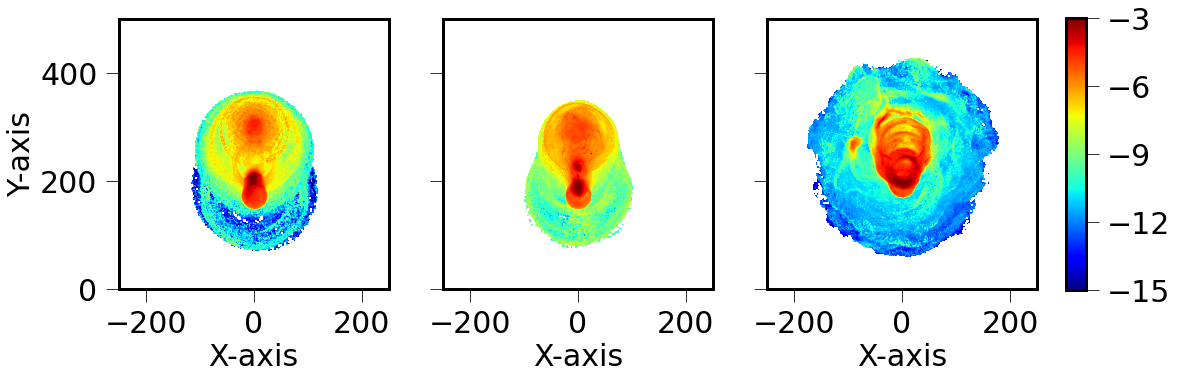}
    
    \caption{2D distribution of specific intensity $I_{\nu}$ (in $\log$ scale) in the plane of the sky at frequency $\nu = 1$\,GHz and time $t=50$ (in code units) after integrating the specific emissivity $J_{\nu}$ along the line of sight. 
    Shown are the maps for steady (left), variable (center), and precessing (right) jets with inclination $\hat{\textit{\textbf{n}}}_{\rm los} = 90^\circ$ (top panel), $45^\circ$ (middle panel) and $10^\circ$ (bottom panel) respectively. 
    The colorbars indicate the level of specific intensity in $\rm{erg \ s^{-1}cm^{-2}str^{-1}Hz^{-1}}$ (in $\log$ scale).
    }
    \label{fig:intensity_1GHZ}
\end{figure*}
%
\subsection{High-resolution Intensity Maps}
\label{sec:hr_maps}
In the following we present the 2D distribution of the specific intensity $I_{\nu}$ at frequency $\nu = 1$\,GHz and time $t=50$ for the different simulations and also for different l.o.s. (see Figure~\ref{fig:intensity_1GHZ}).

In the top panel we show the intensity of the steady, the variable and the precessing jets for an inclination $\hat{\textit{\textbf{n}}}_{\rm los} = 90^\circ$ (i.e. simulation runs \textit{Std90}, \textit{Var90}, and \textit{Prc90}, respectively).
For all three simulations, we find that the intensity maps are highly structured with areas that emit more prominently than others. 
These regions of strongly enhanced local emission we may identify with so-called jet  {\em knots} as they are found in jet observations.

Specifically, we find for the steady jet simulation observed along a l.o.s. of $90^{\circ}$ and for a frequency of $\nu = 1$\,GHz, that the emission is particularly intense at $Y \simeq 200$, which is exactly the site of the \textit{recollimation} shock seen in the MHD simulation.
Also at $Y \simeq 400$, which is the site of the \textit{strong steady}\footnote{The strong steady shock refers to the special feature we found in \citetalias{dubey2023} characterized by very low speed, high compression ratio and very efficient particle acceleration.} shock, we see strong emission. Similarly at $Y \simeq 700-800$, which is the site of \textit{Mach} shock and the \textit{termination} shock. 

The presence and intensity of these features in the steady jet also depends on the magnetic field applied. 
We find that reducing the magnetic field $B_c$ along the jet axis by a factor $100$ leads to a less pronounced recollimation shock. 
This results from the weakening of the toroidal magnetic field, which is interlinked to $B_c$ by the injection profiles we choose \citepalias{dubey2023}, leading to a weaker collimation of the jet.    
In addition, the prominence of the strong steady shock increases.

When observing the variable jet {\em Var90} along a l.o.s. of $90^{\circ}$ and for a frequency of $\nu = 1$\,GHz, 
we see the formation of an elongated knot near the base of the jet, at $Y \simeq 100$. 
This elongated structure can be interpreted as {\em two} {"}bulb{"}-like sub-structures, located at $Y \simeq 100$ and $180$, 
and can be understood as resulting emission from two different bow shocks. 
These bow shocks are formed when the faster jet material interacts with the slower moving material ahead, as induced from the 
variable nature of the jet velocity. 

Interestingly, a smooth, wiggling feature can be seen at $Y\simeq 200$, however, it is very small in size extending just for $\simeq 20$ pixels. 
Further downstream, we see another bulb-like knot from a bow shock at $Y\simeq 320$, followed by an even smaller wiggling feature. 
Additionally, emission from the termination shock region for $Y>600$ is prominent as well. 

We find that the wiggling features mentioned above are more pronounced at lower frequencies and are unique to the variable jet simulations. 
They may be a consequence of a hypothetical streaming instability, or the helical geometry of the magnetic field on these scales in the variable jet. 
Note that these features are found along the jet axis.
It also looks like it would be squeezed out of the divergence of the recollimation shock.
A detailed study of this feature would be interesting, however it is outside the scope of our present work. 

For the precessing jet, seen from a l.o.s. of  $90^{\circ}$, we clearly detect a helical central spine jet - the signature of precession.
In particular, at the jet head we observe multiple knots formed from interaction of the jet head, whose position varies with time as a result of precession, with the ambient medium.

When these jets are seen from aside, with a l.o.s. of $45^{\circ}$, the intensity of the features discussed above is even more enhanced - simply as a result of Doppler boosting. 
In particular, the jet precession feature is seen more clearly at this angle.

Interestingly, we observe the existence of a few loop structures that were not visible for the perpendicular viewing angle in the precessing jet. 
Close to the jet termination, we see a circular arc, which traces the position of jet head as its position varies over time due to precession. Naturally, the extent of all the jets along the $Y$-axis is reduced due to projection effects. 

For a viewing angle of $10^{\circ}$, the linear extension of the jets is further reduced. 
We now see {\em composite} knots representing the combined integrated emission from multiple {\em individual} knots in the jet.
At this l.o.s. it is difficult to investigate different regions of the jet, and we essentially see the core emission from the 
entirety of the jet.
We note here the advantage of our numerical approach compared to an observation, namely that we are able to disentangle the superimposed emission features.
\begin{figure*}
    \includegraphics[width=0.9\linewidth]{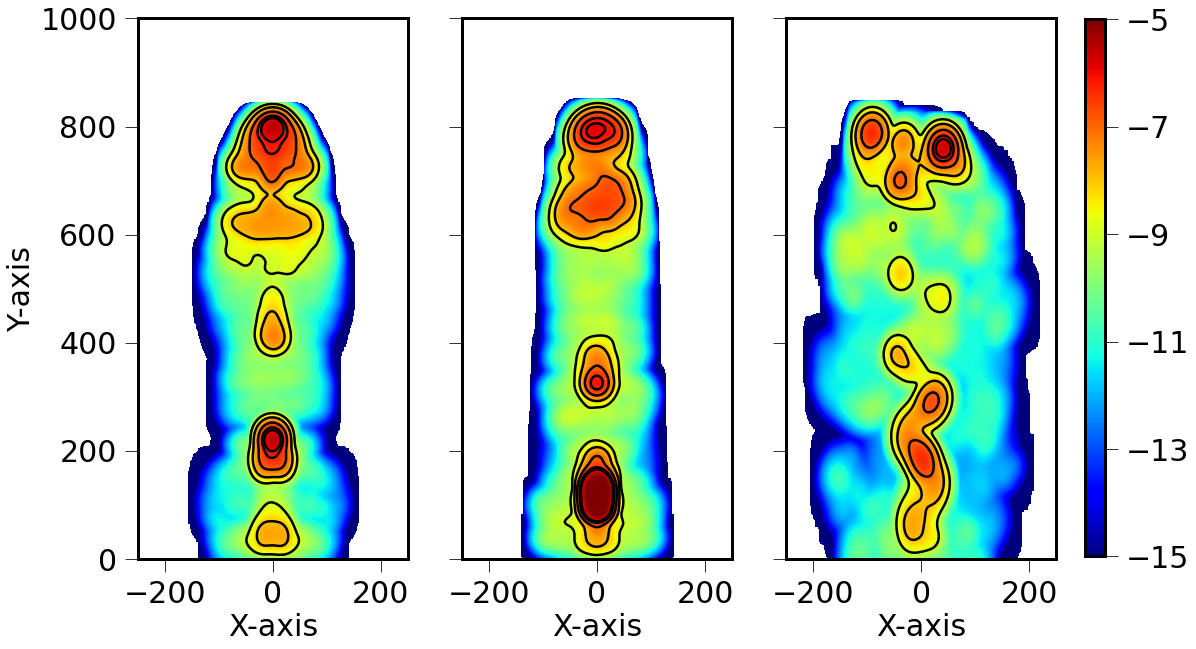}
    \includegraphics[width=0.9\linewidth]{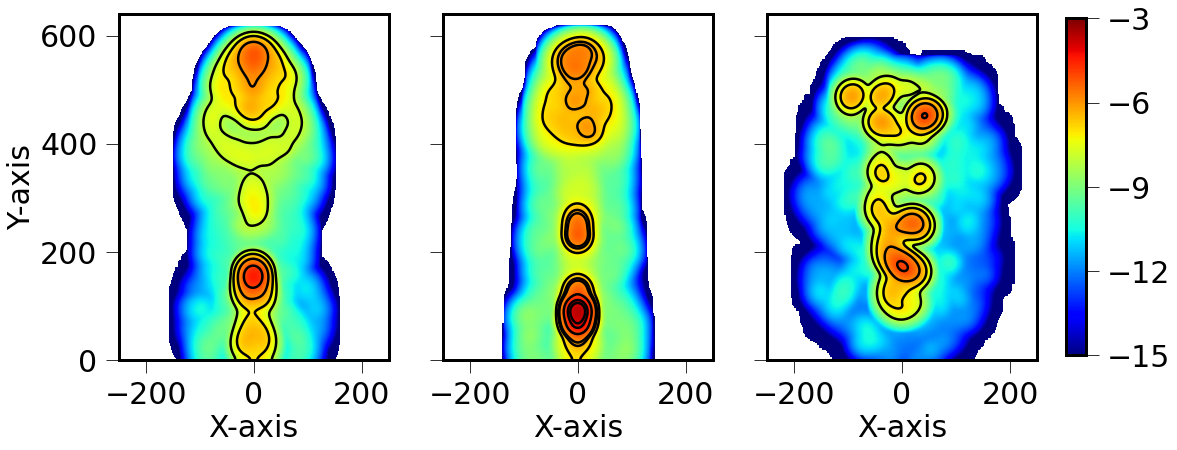}
    \includegraphics[width=0.9\linewidth]{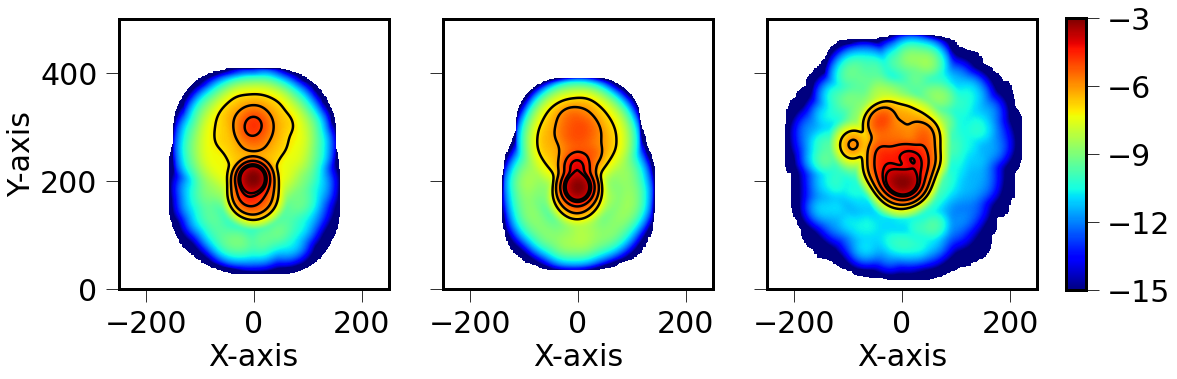}
    \caption{2D distribution of specific intensity $I_{\nu}$ (in $\log$ scale), blurred with a Gaussian beam, in the plane of the sky at frequency $\nu = 1$\,GHz and time $t=50$ (in code units) 
    after integrating the specific emissivity $J_{\nu}$ along the line of sight. 
    Shown are the maps for  
    steady (left), variable (center), and precessing (right) jets with inclination $\hat{\textit{\textbf{n}}}_{\rm los} = 90^\circ$ (top panel), $45^\circ$ (middle panel) and $10^\circ$ (bottom panel), respectively. 
    The colorbars indicate the level of specific intensity in $\rm{erg \ s^{-1}cm^{-2}str^{-1}Hz^{-1}}$ (in $\log$ scale).
    The black contours indicate 0.01, 0.1, 1, 5, and 10 per cent of maximum intensity in all panels. 
    Additionally, contours at $10^{-5}$ of the maximum intensity are added in the middle panel 
    for better visualisation.
    }
    \label{fig:blur_intensity_1GHZ}
\end{figure*}
%
\subsection{Mock Intensity Maps}
\label{sec:blurred_maps}
In order to compare our high-resolution {\em pure} intensity maps, derived directly from the numerical results, with typical observed intensity maps, we need to take into account the finite observational resolution and/or the beam width of the telescope.  

We have therefore convolved the numerical intensity maps with a Gaussian kernel applying a standard deviation of $\sigma = 10$ (simulation cells) in both $X-$ and $Y-$direction. 
This Gaussian kernel is thought to represent a circular telescope beam with a width of $0.2$\,pc, applying the astrophysical scaling of our simulations (see above).
This would correspond to $2.58$\,mas for the case of M87.

We show the post-convolution {\em mock} intensity maps in Figure~\ref{fig:blur_intensity_1GHZ}. 

The existence and shape of the knots is more prominently visible in the mock images, as the {numerically derived features} have now smoothened over the beam. 

The knots in the steady and the variable jets are distributed mainly along the $Y-$axis, whereas in the precessing jet knots are more widely distributed across the jet. 
This is obviously a result of the underlying dynamics of these jets. 
We also see that the overall number of knots in the precessing jet in more than that in the steady and the precessing jet. 
This is directly related to more number of shocks present in the precessing jet as compared to other two jets \citepalias{dubey2023}.
Ultimately, this results from enhanced interaction of the injected material in the precessing jet with other jet material as well as the ambient medium - a consequence of constantly changing direction of injection over time.

The steady jet and the variable jet showing multiple knots as typical for the observed sub-structure of AGN jets \citep{hardcastle2003, cheung2007,giovannini2018}. 
These knots can be either quasi-stationary, stationary, or move with sub-luminal, luminal or even superluminal speeds.
Here, we define the {\em knots} as patterns of localized enhanced emission in the plane of the sky.
We discuss about the speed of these knots in Section~\ref{sec:pattern_speed} in detail.

The knots in the precessing jet seem to follow curved trajectories. 
Such an appearance eventually results from the underlying jet dynamics - the injection along a precession cone and subsequent excitation of shocks - together with radiative features - i.e. the radiation pattern that is generated by shock-accelerated particles that cool down.
These jets when observed from a different l.o.s. may show an $S$- or $Z$-shaped morphology in the radio emission, and may responsible for forming a class of so called {\em winged radio sources} \citep{escamilla2015,yao2021,giris2022,fellenberg2023}. 

\subsection{Multi-frequency Mock Intensity Maps}
\label{sec:multi_freq}
\begin{figure*}
    \centering
    \includegraphics[width=\linewidth]{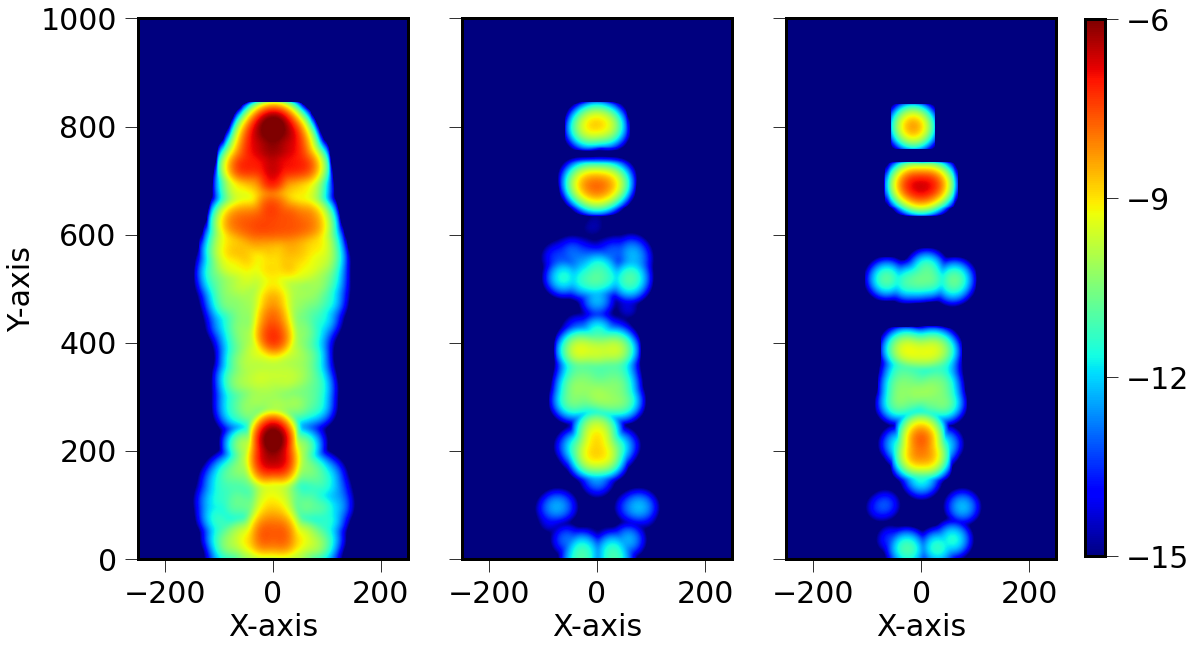}
    \caption{Mock radiant intensity ($\nu I_{\nu}$) distribution (in $\log$ scale) in the plane of the sky  for the steady jet {\em Std90} at time $t=50$ 
    and frequency $\nu = 1$\,GHz (left), $9\times 10^4$\,GHz (center), and $10^9$\,GHz (right), lying in the radio, optical and X-ray regime, respectively. 
    The colorbars indicate the levels of radiant intensity in $10^9 \rm{erg\,s^{-1}cm^{-2}str^{-1}}$ (in $\log$ scale). }
    \label{fig:multi_int}
\end{figure*}

We now investigate from where within the jet the emission at different frequencies originates.
In Figure~\ref{fig:multi_int} we display the distribution of the mock radiant intensity $\nu I_{\nu}$ (where $I_{\nu}$ is the spectral intensity) at different frequencies in the plane of the sky for the steady jet {\em Std90} with l.o.s. at $90^{\circ}$ from the jet axis at $t=50$. 
We choose to show the radiant intensity instead of the spectral intensity in order to account for the large difference in observing frequencies here. 
The radiant intensity is also more closely connected to the luminosity of the jet, which also takes into account the frequency of the observation. 

We see that the radiant intensity $\nu I_{\nu}$ is more pronounced at $\nu =1$\,GHz as compared to other higher frequencies. 
Interestingly, we find that the knot pattern in the radio and optical band at the identical time are not located exactly at the same position, but are slightly {\em shifted} relative to each other. 
In fact, observations of AGN jets show that the position of the {\em core} depends on the observing frequency \citep{blandford1979, fromm2015}. 
As the observing frequency becomes higher, the position of the core shifts closer to the jet base \citep{sokolovsky2011, pushkarev2012, plavin2019}. 

In our {\em synthetic} observations we find a shifting of specific knot patterns as well. 
However, the degree of this shift, as well its dependence on frequency, is not the same for all the knots. 

Essentially, the position of a knot at different frequencies in our simulations depends on the nature of the shock a macro-particle encounters, and its subsequent evolution and cooling. 
Therefore, for example, a macro-particle moving in the region where $Y>600$ first encounters multiple weaker shocks, and is accelerated to relatively moderate $\gamma_{\rm max}$. 
The electrons of this macro-particle subsequently cool down and radiate in the radio band as shown. 
However, moving a little bit more downstream, this particle will encounter the Mach shock which accelerates the electrons to very high $\gamma_{\rm max} \simeq 10^8$. 
Overall, the high-energy electrons in this particle will cool faster to produce optical, and even X-ray, emission as we see. 

At later times, lower-energy electrons, thus with larger cooling time, will radiate predominantly in the radio band. 

Overall, in this particular example, this results in a reduced radio emission at the location of optical and X-ray knots at $Y\simeq 700$, and vice-versa.  

\section{Intensity Evolution \& Pattern Motion}
\label{sec:int_evolution}
Astrophysical jets are not observed as a smooth structure, but are structured in so-called {\em knots}.
These knots are observed as patterns of high emission in the jets. 
Some of them move at seemingly high, even superluminal, velocities, while others appear as stationary. 
After all scientific efforts, today, the exact nature and origin of these knots is not really understood. 

While these knots appear as an emission pattern, the question arises whether and how they are related to dynamical features of the jet, such as regions of certain (particle) density, magnetic field strength? 
Or can these be just {\em patterns} of emission that is unrelated to the dynamics of the underlying jet material?

Our unique approach of applying a time-dependent MHD simulations in order to produce mock intensity maps allows us to investigate the time evolution of the intensity distribution. 
The combined treatment of gas dynamics and emission also allows us to investigate the inter-relation between the radiation pattern and the geometry of the jet material in detail.

In order to investigate these questions, here we study the dynamics of the knots and compare it with the dynamics of the jet material in our simulations.
We first localise these knots in different jets in Section~\ref{sec:knot_localisation}. 
Then, we discuss the evolution and pattern speed of these knots in Section~\ref{sec:pattern_speed}. 
Finally, in Section~\ref{sec:superluminal_motion} we discuss the possibility of superluminal motion in  our synthetic observations.

\subsection{Localisation of Jet Knots}
\label{sec:knot_localisation}
In a first step, we need to locate these knots quantitatively in the radiation pattern of the intensity maps. 
Doing this is comparatively easier for the steady and the variable jets as a result of their linear structure along the jet axis. 
However, in the precessing jet, the knots are scattered across the jet owing to the time-dependent motion of the jet nozzle. 
This makes it difficult to localise knots in this jet. 

For localising the knots in the steady and variable jet simulations at a particular time step, we plot the variation of 2D 
intensity in the plane of the sky along the jet axis at that time. 
We then define knots as the local maxima of the 2D intensity distribution.
Thereafter, the locations of different knots can be defined as the positions of these local maxima of the intensity distribution 
along the jet axis.

\subsection{Intensity Evolution and Pattern Speed of Jet Knots}
\label{sec:pattern_speed}
Once we have identified the knot positions, we can derive their {\em pattern} speed {\em i.e.} the speed of the knot pattern projected in the 2D plane of the sky.
We will discuss the intensity evolution and pattern speed of selective knots for each jet simulation separately in the following sections.

\subsubsection{The Steady Jet Knots }
\label{sec:evol_steady}
\begin{figure*}
    \centering
    \includegraphics[width=\linewidth]{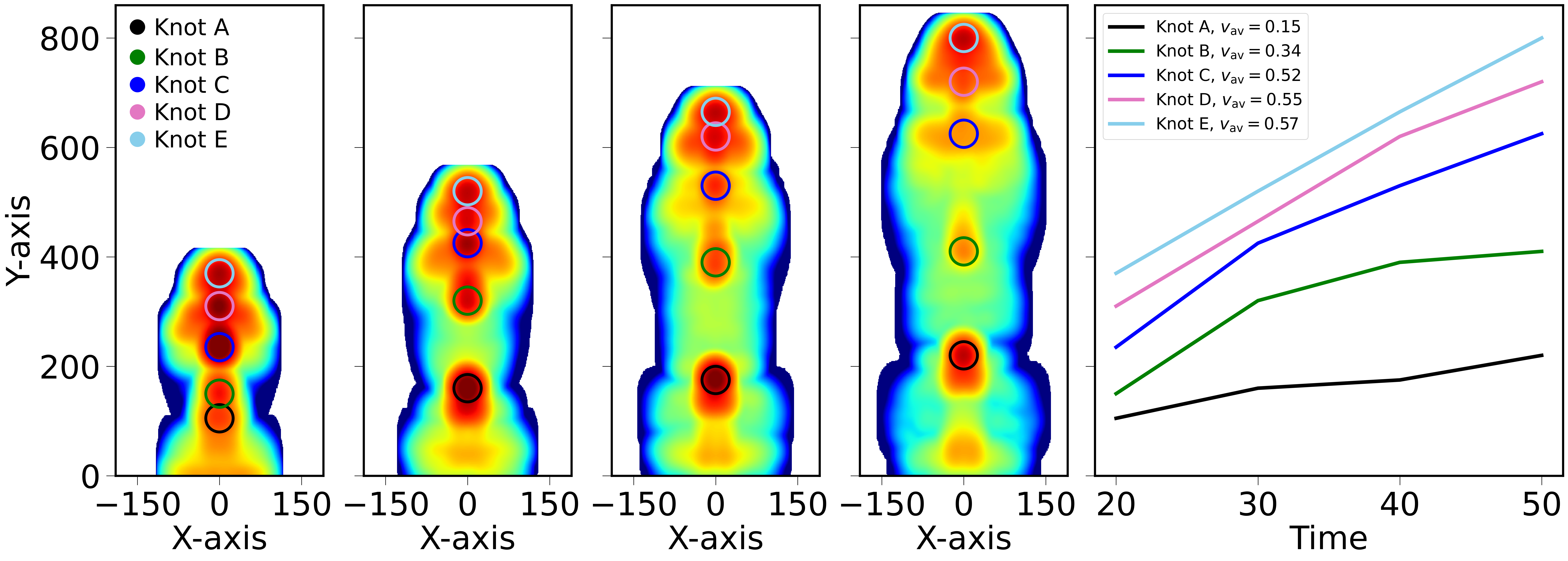}
    \caption{Intensity $I_{\nu}$ map for frequency $\nu =1$\,GHz in the 2D plane of the sky for the steady jet {\em Std90} for times 
    $t = $ 20, 30, 40, and 50 (from left to right). 
    Indicated with different colors are the positions of various knots at different times ({\em left}) 
    and their average pattern speed $v_{\rm av}$ ({\em right}).}
    \label{fig:knotanalysis_Std90}
\end{figure*}
In Figure~\ref{fig:knotanalysis_Std90} we show the evolution of intensity distribution for the steady jet {\em Std90} at $\nu = 1$\,GHz. 
We also indicate the positions of various knots, {\em i.e.} Knots A-E, at times $t = 20, 30, 40, 50$ and also their average pattern speed.

We find that the pattern speed of the Knot E located at the jet head remains fairly constant with an average velocity $v_{\rm av} = 0.57c$. 
This can be simply interpreted as the {\em bulk} velocity of the jet structure, with which it propagates into the ambient medium. 

We see that although we inject the jet with relativistic speed ($\Gamma = 10$), the bulk velocity thus obtained is lower. 
This can be explained as a result of resistance due to the denser ambient medium and expansion of the jet in other directions. 
Knot D, representing another component in the termination shock region has a similar speed as Knot E for the initial times. 
However, it slows down slightly leading to an average pattern speed $v_{\rm av} = 0.55c$.
We note here that the pattern speed of Knots D and E are similar also to the propagation speed of termination bow shock, 
which moves with a speed $\simeq 0.5c$.

Knot C, on the other hand, is resulting from the Mach shock and moves with an average pattern speed $v_{\rm av} = 0.52c$. 
However, this pattern speed is time dependent and is quite faster between time $t=20-30$, with $v_{\rm av} = 0.76c$. 

Knot B is related to the strong steady shock, and has an average pattern speed $v_{\rm av} = 0.34c$. 
This knot too, like Knot C, has a higher initial speed $v_{\rm av} = 0.68c$ between time $t=20-30$. 
Compared to the pattern speed, the speed of the strong steady shock itself is only $0.1c$. 

Hence, although the emission from a knot (Knot B here) is governed by the shock (strong steady shock here), 
the dynamics of the knot and the corresponding shock need not to be the same and may differ.
Essentially, the particles - though accelerated by the shock - may, while cooling down, travel away from the shock with a speed that is different from the shock speed. 

Essentially, although the electrons in a macro-particle are accelerated by shock, while cooling down the macro-particle may travel away from the shock at a speed that is different from the shock velocity.
In our approach, this speed of the (macro-)particles is the same as the fluid speed.

The Knot A is associated with the recollimation shock, and has an average pattern speed of $v_{\rm av} = 0.15c$. 
Its speed does not fluctuate much with time, and hence it may be considered as a stationary feature.

By comparing the macro-particle composition of various knots at different times, we also find that at the different times these knots are composed largely from different macro-particles, even though they may be stationary.
Hence, there is a constant flux of macro-particles through the knots, even though the knot may remain stationary. 
This further suggests that knots are essentially {\em patterns} of emission, and are connected to the shocks rather than dynamical flow of radiating particles or fluid. 

\subsubsection{The Variable Jet Knots}
\label{sec:evol_variable}
\begin{figure*}
    \centering
    \includegraphics[width=\linewidth]{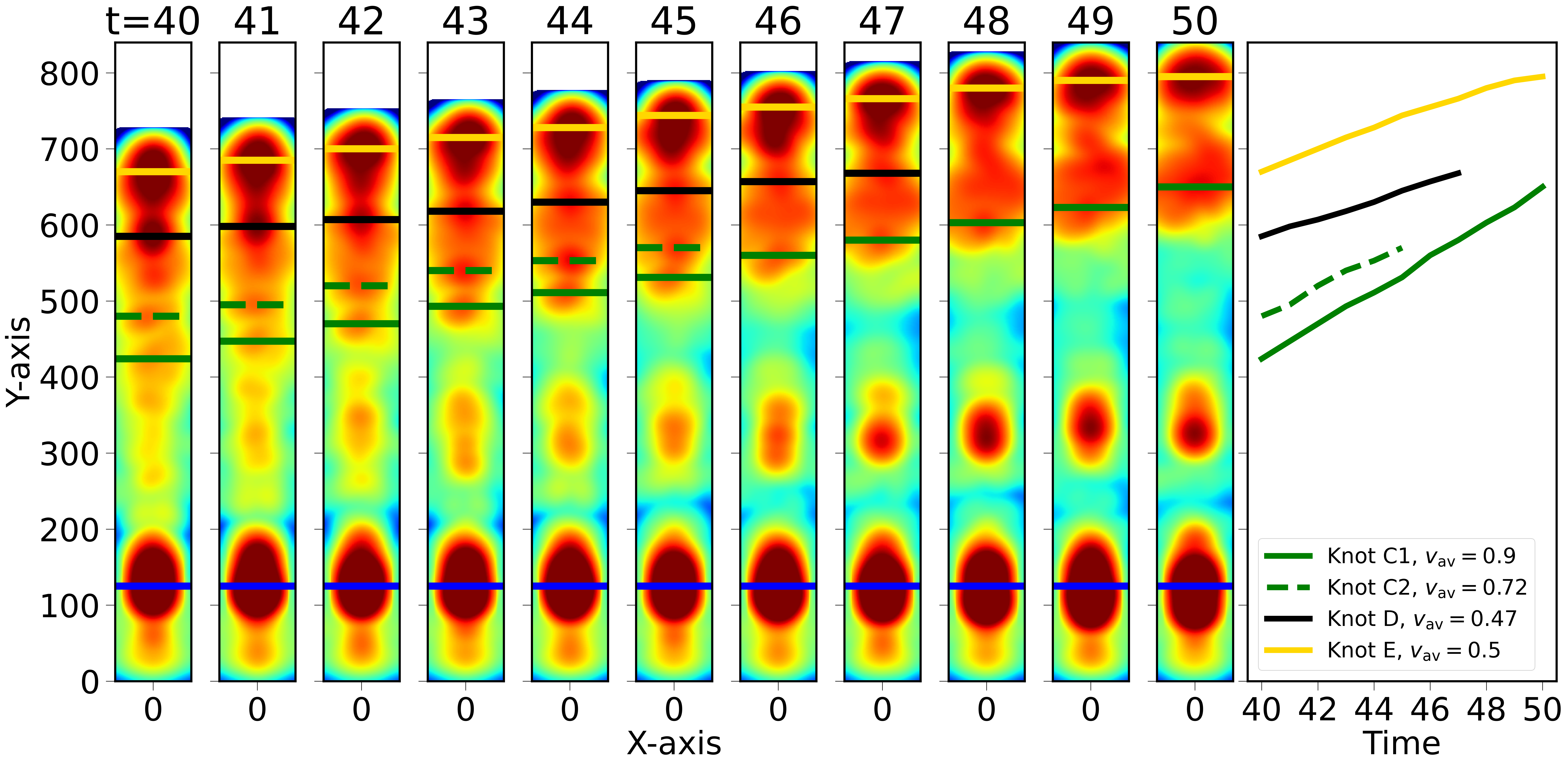}
    \caption{Intensity distribution $I_{\nu}$ at $\nu =1$\,GHz in the 2D plane of the sky for the variable jet {\em Var90} for $t = 40-50$, 
    as mentioned in the panel titles. 
    Indicated, with different colors, are the positions of various knots and their average pattern speed $v_{\rm av}$.}
    \label{fig:knotanalysis_VarNorm90}
\end{figure*}

The variable jet {\em Var90}, on the other hand, has a much more complex structure as compared to the steady jet regarding emission features. 
This is simply a result of the more complex underlying dynamics governed by the time-variable injection. 
Hence, we choose to study its intensity evolution in a much more detail. 
In Figure~\ref{fig:knotanalysis_VarNorm90}, we show the evolution of 2D intensity distribution between time $t=40-50$ for the variable jet {\em Var90} at $\nu = 1$\,GHz. 
Again we indicate the position of various knots at different times through different colors. 
Note that here, we focus just on the structure close to the $Y-$axis where the knots are located. 

We find that the knots in the variable jet are more variable than the steady jet as these knots show fading and flaring, as well as ejecting patterns from them and merging with other knots. 

Knot A (blue color) can be considered as a stationary knot since its position does not change at all during this period. 
This may, in principle, be a proxy for observed stationary knots, such as {\em e.g.} HST-1 in M87 \citep{cheung2007, nakamura2010}, or knot C7 in BL Lac \citep{cohen2014}.

Going further downstream, at $Y\simeq 300$ we identify Knot B which varies rapidly with time. 
At $t=40$, this irregularly shaped knot looks fragmented and has comparably lower intensity. 

As time evolves, the knot appears to consolidate, as though from merging of various fragments, and forms a more regular shape
at time $t=48$. 
During this process the knot also flares up to higher values of intensity, increasing from $I_{\nu} \simeq 10^{-8} \ \rm{erg \ s^{-1}cm^{-2}str^{-1}Hz^{-1}}$ at $t = 40$ to $I_{\nu} \simeq 10^{-6} \ \rm{erg \ s^{-1}cm^{-2}str^{-1}Hz^{-1}}$ at $t = 48$.
This may result from the merging of various bow shocks, moving at different velocities and formed as a result of variable jet injection. 
The resultant shock becomes much stronger leading to enhanced particle acceleration. 
Subsequent cooling becomes more efficient and leads to stronger radiation. 

Knot C at $Y\simeq 650$ at $t=50$, seems to be composed by a merger of two components -- Knot C1 (shown by solid green line) moving with an average speed $v_{\rm av} = 0.9c$, and Knot C2 (shown by broken green line) moving with an average speed $v_{\rm av} = 0.72c$. 
These two components merge at time $t=47$ to form a single knot, which then evolve as a single knot. 

Knot D (shown by black line), which has a rather higher intensity at $t=40$, starts to dim fade thereafter. 
Moving with an average speed of $v_{\rm av} = 0.47c$, it loses is integrity at time $t=48$, after which we can not disentangle any individual components. 

Lastly, Knot E (shown by yellow line), moving with an average speed of $v_{\rm av} = 0.5c$, results from the termination shock that is formed by the interaction of the jet head with the ambient medium. 

\subsubsection{The Precessing Jet Knots}
\label{sec:evol_precessing}
\begin{figure*}
    \centering
    \includegraphics[width=\linewidth]{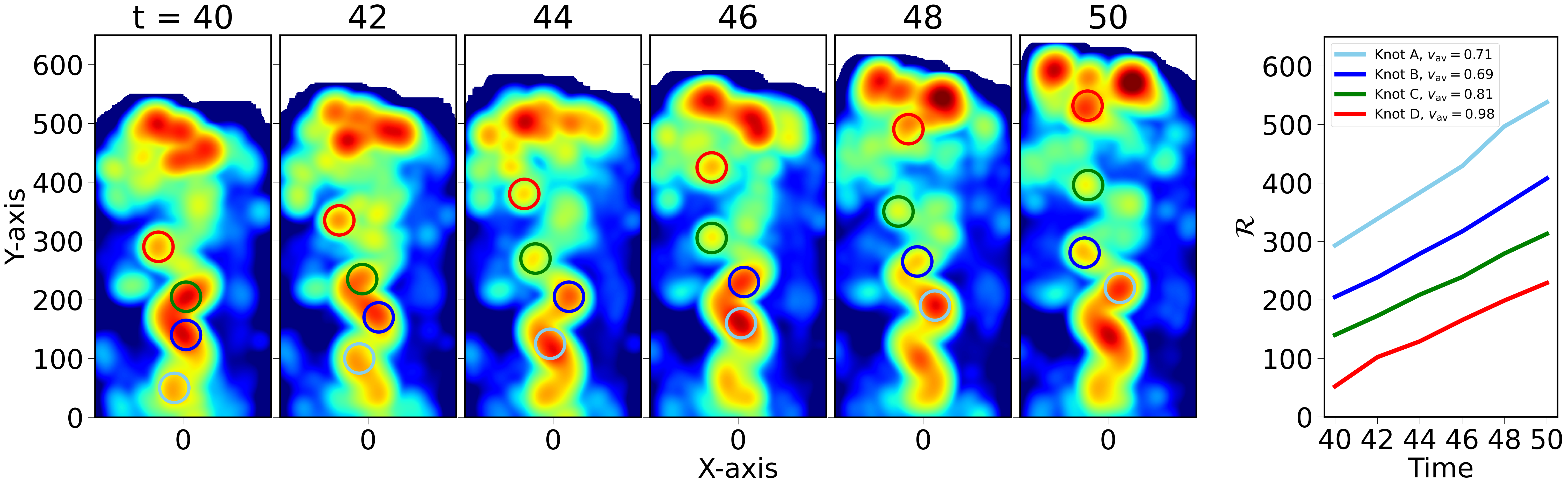}
    \caption{Intensity distribution $I_{\nu}$ at $\nu =1$\,GHz in the 2D plane of the sky for the precessing jet simulation {\em Prc90} for times $t = 40-50$, 
    as mentioned in the panel titles. 
    Shown also, with the respective colors, are the relative distance $\mathcal{R}$ of various knots, from which we calculate their average pattern 
    speed $v_{\rm av}$.}
    \label{fig:knotanalysis_Prc90}
\end{figure*}
The knots in the precessing jet are harder to identify and localize. 
This is a consequence of the directional change of the jet nozzle with time, leading to a curved morphology and off-axis knots. 
Thus, a study of the evolution of knots, in this case, is impacted highly by which features we identify as belonging to the same knot in the course of its evolution. 
This limitation of correctly identifying the features belonging to the same knot at different times is typically known from observational studies of the jets as well.

As previously mentioned, in the precessing jet, different to the steady and the variable jet, we see knots also moving across the jet, in addition to moving along the jet axis. 
Consequently, in order to quantify the pattern speed of the knots, we need to define properly the distance travelled by a knot in a period of time. 

We define the position of a knot in the $X-Y$ plane at time $t_i =42$ as ($X_i, Y_i$). 
The distance of this knot from the origin $X=Y=0$ is then given by $\mathcal{R}_i = \sqrt{X_i^2 + Y_i^2}$. 
At a later time $t$, this knot has moved to position ($X_t, Y_t$) at a distance $\Delta \mathcal{R}$ from the initial position. 
We define the distance of the knot at time $t$ as $\mathcal{R} = \mathcal{R}_i + \Delta \mathcal{R} = \mathcal{R}_i + \sqrt{(X_t - X_i)^2 + (Y_t - Y_i)^2}$.

In Figure~\ref{fig:knotanalysis_Prc90}, we show the evolution of the intensity distribution for the precessing jet
{\em Prc90} at $\nu = 1$\,GHz for the times $t = 40-50$.  
We also show the relative distance $\mathcal{R}$ the various knots move as a function of time and also their average pattern speed $v_{\rm av}$. 

As a general feature, we see enhanced emission from those regions where the main jet path turns along the helical precession pattern. 
This can be explained simply as a result of integrating the light along l.o.s. which here is a tangent to the precessing cone, 
where we find a larger number of active shocks, with high emission similar to limb-brightening.  

In the region close to jet termination we find a highly complex pattern of a number of knots.  
This is a consequence of interaction of moving jet head with the ambient medium. 
As the jet head changes its position with time, various locations in the region become the sites of shocks and thus of emission.
Subsequently, these knots cool and fade when the jet head moves further downstream and away from that shock locations.
As seen previously also for the steady jet and the variable jet, the speed of the jet head - hence of the (macro-)particles there - and the termination shock is similar in the precessing jet as well. 
As a result, the pattern speed of the knots corresponding to the termination shock move with similar speed as the termination shock and the jet head.

We now briefly discuss certain interesting knots in this jet. 
Upstream from the jet termination, we have define  Knot D (in red color). 
This knot moves with a high average velocity $\simeq 0.98c$ until it reaches the termination shock, and is then integrated as part of complicated knot structure there. 

Further upstream, Knot C (in green color) moves with a lower average pattern speed $\simeq 0.81c$ on a seemingly helical trajectory. 
This helical trajectory is a result of precession of jet spine, and it more visible and sustained in the region closer to the jet base.
Note that the jet material is {\em not} moving on a helical trajectory, as it is injected at a particular instance into a {\em ballistic trajectory}.

With the jet time evolution, the precession cone further opens up and expands as well in lateral direction.
As a result,  Knot B (in blue color) and Knot A (in sky blue color) show an enhanced motion also in the $X$-direction\footnote{Note that this lateral motion of knots 
cannot be found for the steady and the variable jet dynamics.}, as compared to Knots C and D.
The Knots A and B move with an average pattern speed $\simeq 0.71c$ and $\simeq 0.69c$, respectively.

As the jet expands, it cools down as well. 
This leads to weakening of the shocks, and consequently in a general fading of the knots as they move downstream, {\em e.g.} Knot C, which has intensity $I_{\nu} \simeq 3 \times 10^{-7} \ \rm{erg \ s^{-1}cm^{-2}str^{-1}Hz^{-1}}$ at $t = 40$, fades to $I_{\nu} \simeq 7 \times 10^{-9} \ \rm{erg \ s^{-1}cm^{-2}str^{-1}Hz^{-1}}$ at $t = 50$.

In general, we see that the pattern speed of knots in the precessing jet is in fact the highest of all three jets we investigated, followed by the variable jet and the steady jet. 
\\
\\
Overall, these results suggests that a time-dependent injection of jet material in general leads to a higher pattern speed of the knots as compared to a constant, time-independent injection. 
Hence, we may hypothesize that observed knots showing superluminal motion may be generally related to a time-dependent {\em outburst} of jet material.

\subsection{Superluminal Motion}
\label{sec:superluminal_motion}
Radio observations of relativistic AGN jets have revealed jet knots that seem to travel with superluminal speed. 
In fact, observed knot pattern have been observed reaching projected speeds of even $\sim 40c$, in particular at parsec scales,
although commonly they show speeds between $1-10c$ \citep{lister2016, kim2018, giovannini2018, walker2018m87}. 

Superluminal motion simply results from the effects of projection combined with light travel time in case of a relativistic motion of a radiation source towards the observer.
Moving with high speed $v\simeq c$, and with angle $\theta$ between the velocity vector of the radiation source and the l.o.s.,
the apparent velocity $v_{\rm app}$ of such source in the plane of the sky is $ v_{\rm app} = v \sin{\theta}/\left(1-v\cos{\theta}\right)$. 

In the previous section, we discussed the pattern speed of selected knots in the different jet models we investigate, for a l.o.s. of $90^{\circ}$. 
We found that many knots move at sub-luminal average pattern speeds.
However, for certain duration of their evolution specific knots can have a highly relativistic pattern speeds close to the speed of light.
In particular, Knot C in the steady jet {\em Std90} between time $t=20-30$ moves with a pattern speed of $0.76c$. 
Similarly, Knots C1 and C2 in the variable jet {\em Var90} move with average pattern speed of $0.9c$ and $0.72c$, respectively. 

With the derived pattern speed, we can now tell whether this pattern motion would result in an apparent superluminal speed when viewed from a l.o.s. closer to the jet axis. 
We find that for a speed $v=0.72c$ corresponding to average pattern speed of Knot C2 in the variable jet, we can get a maximum apparent speed $v_{\rm app} \simeq 1.1c$ at a l.o.s. angle $\theta \simeq 45^{\circ}$. 
Knot C in the steady jet, on the other hand, moving with a pattern speed $v = 0.76c$ between $t=20-30$, may be observed with a maximum apparent speed $v_{\rm app} \simeq 1.2c$ at a l.o.s. angle $\theta \simeq 40^{\circ}$. 
Therefore, these patterns would move superluminally only marginally.

On the other hand, Knot C1 in the variable jet shows a highly relativistic average pattern speed $v = 0.9c$. 
When viewed from $\theta \simeq 25^{\circ}$, its apparent motion would be $v_{\rm app} \simeq 2c$. 

The knots in the precessing jet move with the highest pattern speeds of all the jets we have studied. 
Knot D in the precessing jet moves highly relativistically with an average pattern speed $\simeq 0.98c$. 
This knot, when viewed from an l.o.s. $\theta \simeq 10^{\circ}$, would move with $v_{\rm app} \simeq 5c$.

While observed radio knots may move with several tens of c, in our simulations we get the highest possible apparent speed of $\sim 5c$. 
We believe that this is due to the relatively smooth and dense jet structure, in spite of all the variable dynamics present.
We therefore conjecture that jet simulations on a larger grid and with higher amplitudes in mass flux may result in a higher pattern speed of their knots.
In this case, a previous jet injection will clear the way for the next injection, thus allowing for higher differential speed and thus higher shock speeds.
The following jet will just find a domain with lighter jet material, through which the knots may move faster. 

We emphasize that the derivation of the knot apparent velocity depends highly on what features we consider as belonging to the same knot at different times. 
This decision can be made more consistently through simulation, like ours, as a result of high resolution and availability of emission data for each particle. 
This problem is well known from observational studies. Indeed, in observations, it may be more difficult to recognise knot correlations because of the lower resolution as well as the lower sensitivity of the telescopes. 

To summarize this section, we have found that superluminal motion can be observed for all three variations of jets we investigate. 
However, the maximum apparent pattern speed the jet can produce depends on the dynamics of the jet. 
In the steady jet, we find possibility of only a marginal superluminal motion with apparent speed close to $c$. 
On the other hand, the time-dependent variable and precessing jets show superluminal apparent pattern speeds up to $\sim 2c$ and $\sim 5c$, respectively.

\section{Synthetic Light Curves}
The emission from relativistic jets is generally observed to be variable on time scales ranging from days to years.
In particular, blazars are known to be highly variable on time scales even as short as minutes \citep{schmidt1963, rani2017, shukla2020, gokus2024}. 
Flux variability may also depend on the frequency, with radio flux sometimes lagging behind the $\gamma-$ray flux \citep{max-moerbeck2014, fuhrmann2014}. 

Our study allows us to study also the variability of jet emission, just by comparing the luminosity emitted by our simulated jets at different times - either by comparing single components or features (i.e. emission from knots), or the radiation emitted by the whole jet structure (i.e. a "core" emission).

\label{sec:light_curves}
\begin{figure*}
    \centering
    \includegraphics[width=\linewidth]{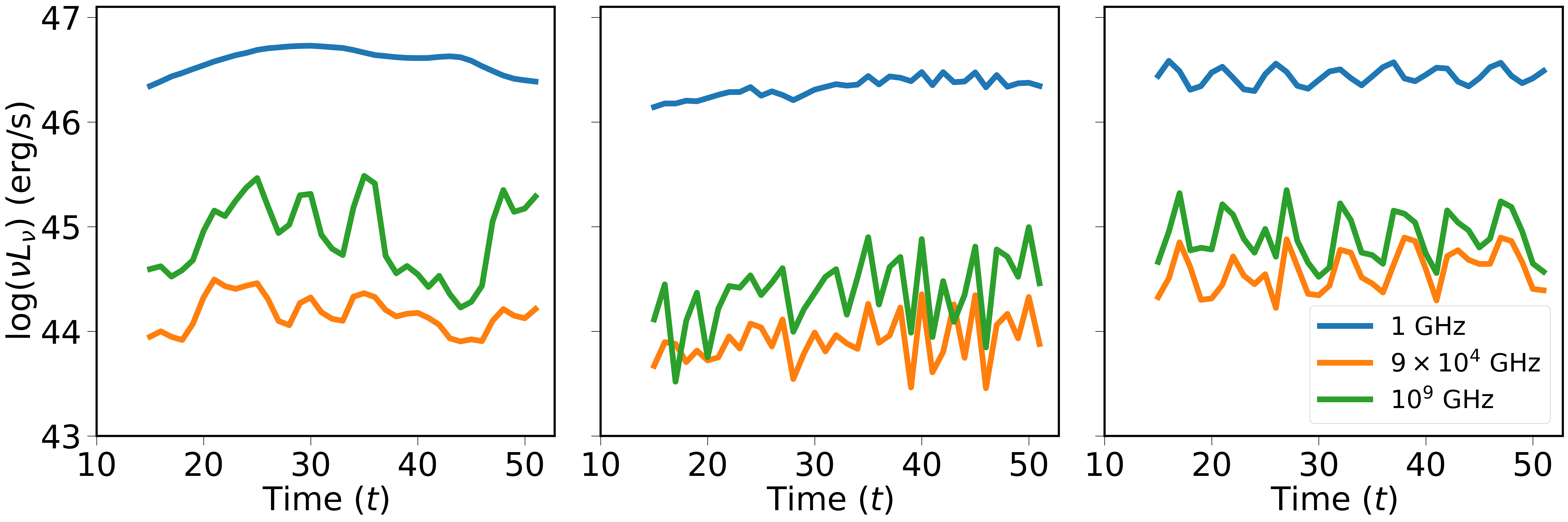}
    \caption{Light curves of the integrated jet luminosity $\nu L_{\nu}$ showing variation of $\log$ of total luminosity with time $t$, 
     representing an unresolved {\em core} luminosity. Shown are the values at frequency $\nu =$ $1$, $9\times 10^4$, and $10^9$\,GHz in blue, orange, and green color respectively, for the jet {\em Std10} (left), {\em Var10} (center), and {\em Prc10} (right).
    }
    \label{fig:lc_norm10}
\end{figure*}

In Figure~\ref{fig:lc_norm10}, we show exemplary {\em core} light curves resulting from the jets generated by our
different jet injection nozzles.
Here, the l.o.s. is inclined by $10^{\circ}$ against the initial jet axis, a geometry more related to blazar jets.
The light curves cover a time scale $t = 5-50$.
We have chosen frequencies of $\nu = 1$, $9\times10^4$, and $10^9$\,GHz lying in the radio, optical, and the X-ray regime, respectively . 

We find that the knot patterns in the jets are dynamic features, and may fade or flare as they evolve (Section~\ref{sec:int_evolution}).
This results in time variability of our {\em synthetic} light curves as well. 
In fact, we find some degree of variability in the core luminosity of all the three jets {\em viz.} the steady, the variable and the precessing jet, and this for all the three frequencies we investigate. 

We also see that, in general, the amplitude of variability of the jet increases as we increase the frequency from radio to X-ray band.
The radio luminosity in all simulated jets fluctuates typically between $0.6-1.4$ times the respective mean luminosities in a period $t=15-50$. 
On the other hand, the optical luminosity in the jets fluctuates somewhat larger, between $0.3-2.4$ times the respective mean luminosities in the same period. 
The variability in the X-ray band is more pronounced, with fluctuations of $0.1-3.2$ times the respective mean luminosities for all jets.
This is a consequence of larger emitting regions in radio band as can be seen from the diffused radio emission (from the cocoon) in Figure~\ref{fig:blur_intensity_1GHZ}. 
In comparison, the emission in the optical and X-ray bands is well localized in knots (Figure~\ref{fig:multi_int}).  
As a result, the fluctuations of luminosity in the radio band are averaged out, while in the optical and X-ray bands are retained in the light curves. 

In the $1$\,GHz radio band, the luminosity of the steady jet {\em Std10} varies only slowly over large time-scales. 
This is a result of a stable dynamics from the steady injection of jet material over time. 
On the other hand, the jets with time-dependent injection {\em viz.} the variable jet and precessing jet, show variability even in the radio band. 
The radio variability is even more pronounced in the precessing jet as the inclination of l.o.s. is the same as the opening angle 
of the precession cone, both being $10^{\circ}$. 
As a result, once in each precession cycle of period $\mathcal{P} =5$, the jet is exactly aligned to the l.o.s., 
leading to Doppler boosting of the luminosity, as evident in the corresponding light curve.

These fluctuations in the radio band may seem marginal, but we note the log scale for the luminosity in the figure,
which is visually suppressing the amplitude of the fluctuations.
While the steady jet luminosity changes smoothly, and only over a long time scale, the fluctuations for the other 
two jet simulations kind of follow the periodicity of the injection nozzle with a period $\mathcal{P} = 5$.

In fact, the ratio $L_{\rm max}/L_{\rm min}$ of the maximum and minimum luminosity at $\nu = 1$\,GHz is $5.6$, $3.2$, and $8$ for the steady, the variable, and the precessing jets, respectively.

Overall we see that in all jets we investigated, the radio luminosity exceeds the luminosities in optical and X-ray bands, whereas the X-ray luminosity exceeds the optical luminosity. 
Note that although the specific luminosity $L_{\nu}$ is lower in the X-ray band as compared to other bands, 
the total luminosity $\nu L_{\nu}$ becomes high as result of very large frequency $\nu$.

Another observational feature in the light curves of the AGN jets is, as mentioned above, the lag between the radio and $\gamma -$ray flux.
However, we find a strong correlation between the light curves at different frequencies with respective correlation coefficients between $0.7-0.8$. 
This can be explained as a result of not considering the finite light travel time between two regions of the jet. 

As we discuss in Section~\ref{sec:multi_freq}, the position of knots at different observing frequencies in our simulations are shifted relative to each other. 
When considering the light travel time, this will lead to the lag actually observed between the light curves at different frequencies. 

\section{Jet spectral energy distribution}
\label{sec:sed}
\begin{figure*}
    \includegraphics[width=\linewidth]{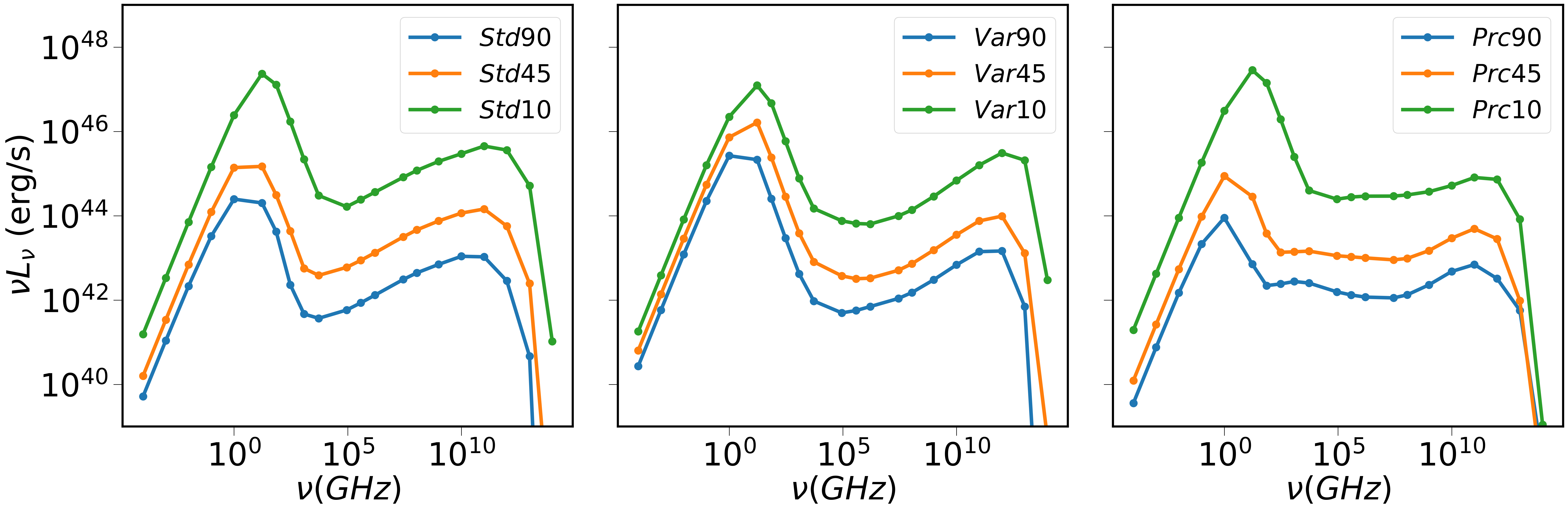}
    \caption{Radiation spectra from simulations of a steady jet (left), a variable jet (center), and a precessing jet (right),
    at time $t=50$ (in code units). 
    In all three panels, the blue, orange and green curves represent the inclination 
    $\hat{\textit{\textbf{n}}}_{\rm los} = 90^\circ$, $45^\circ$, and $10^\circ$ respectively.
    Fluxes are integrated over the whole computational domain, similar to an unresolved {\em core} emission.}
    \label{fig:spc_t50}
\end{figure*}
We now study the dependence of the emitted radiation flux on the frequency $\nu$.
We will show synthetic spectrum of different jets by the total luminosity $\nu L_{\nu}$ (Equation~\ref{eq:flux}). 
We will discuss first the core\footnote{By {\em core} luminosity we denote the integration of intensity (Equation~\ref{eq:intensity}) over the whole domain, implying that the underlying jet is not resolved.} synthetic spectrum produced by the jet in Section~\ref{sec:core_spectrum}. 
We then connect emitted radiation to the underlying particle acceleration and cooling processes in Section~\ref{sec:gmax}. 

\subsection{Core Jet Spectrum}
\label{sec:core_spectrum}
We show the total synchrotron luminosity $\nu L_{\nu}$ (in erg/s) for the simulation setup {\em Std90}, {\em Std45}, and {\em Std10}, respectively, as a function of frequency $\nu$ and for a chosen time $t=50$ in the first panel of Figure~\ref{fig:spc_t50}.

The steady jet simulations result in a classical double-humped SEDs observed in a variety of jet combining the data from multiple wavebands \citep{balokovic2016}. 
The low-frequency peak can be easily explained as a result of synchrotron emission.

However, in the literature the high-frequency peak is discussed either to be caused by IC scattering of the photon field or due to synchrotron emission from a different population of electrons \citep{meyer2011, breiding2017, borse2021}.
Since in this paper we only consider the {\em synchrotron} luminosity for the emission maps, the presence of high-frequency peak (in addition to the low-frequency peak) suggests the presence of multiple populations of electrons in our simulations. 
We discuss the nature and location of these multiple populations of electrons in Section~\ref{sec:population}.

This double-hump feature is also present in the SEDs of the precessing jet (i.e. simulations {\em Prc90}, {\em Prc45}, {\em Prc10}), and in the variable jet (i.e. simulations {\em Var90}, {\em Var45}, {\em Var10}), as shown in Figure~\ref{fig:spc_t50}.

We find a clear increase in the luminosity in all the jet simulations when we decrease the l.o.s. and look into the jet closer to the jet axis.
This can be explained as most of the emission from the jet is strongly beamed (Doppler boosted) in a narrow cone in the direction of propagation. 
However, it is the first time that such effects are self-consistently derived from a relativistic MHD jet simulation.

This increase in luminosity is much more pronounced in the steady and precessing jet simulations as compared to the variable jet. 
This can be explained by the fact that the average bulk Lorentz factor of the variable jet is lower than the steady and precessing jets, and the beaming of luminosity depends on the Lorentz factor. 

In addition to the relativistic boosting, we also observe the Doppler shift of the peak frequencies to higher values when the inclination to the l.o.s. $\hat{\textit{\textbf{n}}}_{\rm los}$ decreases. 

As for the boosting, this shift is less pronounced in the variable jet simulation, again as a result of lower average Lorentz factor and its inverse relation with the corresponding Doppler factor ${\mathcal{D}}$ (Equation~\ref{eq:Doppler_factor}).

We summarize the different values of peak frequency $\nu_p$ and the total luminosity at peak frequency $\nu_p L_{\nu_p}$ for the different simulations in Table~\ref{tab:lum_data} for time $t=50$. 
We find that the peak luminosity $\nu L_{\nu}$ is highest for the variable jet, followed by the precessing jet and the steady jet for inclinations $\hat{\textit{\textbf{n}}}_{\rm los}$ of $90^{\circ}$, $45^{\circ}$, $10^{\circ}$ from the l.o.s., respectively. 
 
The jet spectrum is naturally also governed by the jet magnetic field strength.
We have investigated the impact for the steady jet simulation where we varied $B_c$ along the jet axis. 
We find that reducing $B_c$ by a factor 100 leads to a reduced low-frequency peak luminosity and enhanced high-frequency peak luminosity. 
We discuss more on what causes this change in Section~\ref{sec:shockstructure} where we discuss the spectrum of different jet sections.

In \citetalias{dubey2023}, from our study of the nature of shocks, we found that the shocks in the variable jet were stronger, but less in number, compared to the precessing jets and the steady jets. 
From the electron energy spectrum, we had derived that the particles in the variable jet could be accelerated to lower energy only, as compared to the steady jet and the precessing jet. 

These results suggested that the number of shocks in a jet is more important than the strength of the shocks in regard to {\em particle acceleration}. 
In the present paper, however, we find that at an inclination of $90^\circ$ to the l.o.s. the luminosity of variable jet is 10 times more than the steady and precessing jet at the low-frequency hump with frequency $\nu \sim 1$\,GHz.
Overall, this suggests that the strength of the shocks plays a more significant role than their number in regard to the 
{\em emission} from the jet. 
\begin{deluxetable}{ccc}
\tabletypesize{\small}
\tablewidth{0pt}
\tablecaption{Peak Frequencies and Luminosities \label{tab:lum_data} }
\tablehead{
\colhead{Identifier}  & \colhead{$\nu_{p}$}(GHz) & \colhead{$\nu_{p}L_{\nu_{p}}$(erg/s) }
}
\startdata
\textit{Std90} & $1$, $10^{10}$ & $2.5\times10^{44}$, $1.1\times10^{43}$\\
\textit{Std45} & $17$, $10^{11}$ & $1.5\times10^{45}$, $1.4\times10^{44}$ \\
\textit{Std10} & $17$, $10^{11}$ & $2.3\times10^{47}$,  $4.5\times10^{45}$\\
\textit{Var90} & $1$,$10^{12}$ & $2.6\times10^{45}$, $1.5\times10^{43}$\\
\textit{Var45} & $17$,$10^{12}$ & $1.6\times10^{46}$, $9.8\times10^{43}$\\
\textit{Var10} & $17$,$10^{12}$ & $1.2\times10^{47}$, $3.1\times10^{45}$\\
\textit{Prc90} & $1$,$10^{11}$ & $8.9\times10^{43}$, $6.9\times10^{42}$\\
\textit{Prc45} & $1$, $10^{11}$ & $8.7\times10^{44}$, $4.9\times10^{43}$ \\
\textit{Prc10} & $17$, $10^{11}$ & $2.8\times10^{47}$, $8.1\times10^{44}$ \\
\enddata
\tablecomments{Low and high peak frequencies $\nu_p$ (in GHz) and total luminosity at the peak 
frequencies $\nu_p L_{\nu_p}$ (in erg/s) for various simulation setups as indicated.
}
\end{deluxetable}

Although all the three jets with the different dynamics we discuss, namely the steady, the variable and the precessing jets, 
have double-hump spectra, we find that there are sophisticated differences in their emission spectrum. 
We find that (i) while the difference in luminosity between the low-energy and high-energy peak is large for the variable jet, 
it is less pronounced for the steady and the precessing jet. 

This suggests that the variable jet is much more efficient in radiating at lower frequencies.
We also find that (ii) for frequencies $\nu > 10^{3}$\,GHz, the precessing jet features a flatter spectrum as compared to the steady 
and the variable jet. 
This can be explained by a higher efficiency of the precessing jet for emission at higher frequencies.
We note that this is resulting also in a much more pronounced emission in the X-ray band if compared to the steady and the variable jet. 

\subsection{Particle Acceleration and Emission Spectrum}
\label{sec:gmax}
\begin{figure}
    \includegraphics[width=\linewidth]{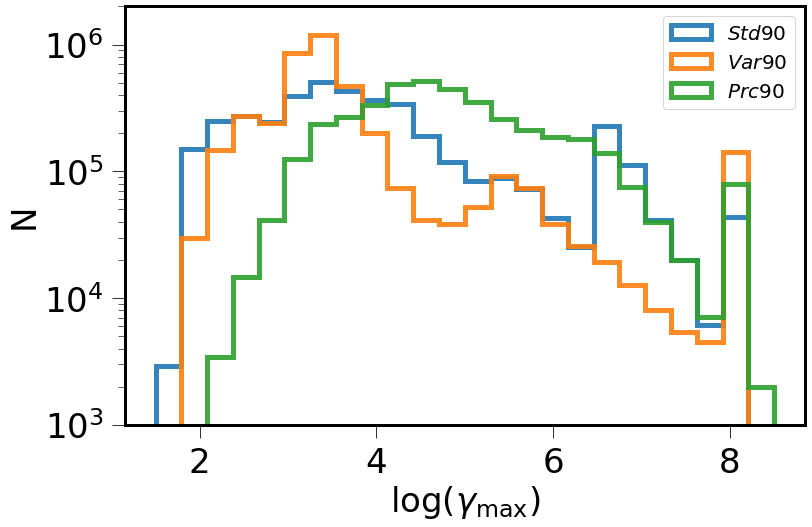}
    \label{fig:hist_gmax}
    \caption{Histogram showing the number of (macro-)particles (in $\log$ scale) with respect to their maximum electron energies $\gamma_{\rm max}$ 
    for simulation {\em Std90} (blue), {\em Var90} (orange), and {\em Prc90} (green) 
    at time $t=50$, accounting for all the particles present in the domain.}
\end{figure}

In order to investigate the different jet emission in more detail, and to connect the emission properties to the particle energy distribution, 
we show in Figure~\ref{fig:hist_gmax} a histogram presenting the number of (macro-)particles versus
the maximum Lorentz factor $\gamma_{\rm max}$ 
of electrons in the energy spectrum of a macro-particle.

We remind that in our approach the actual value of $\gamma_{\rm max}$ of a (macro-)particle  results from the (time-dependent) interplay of cooling and acceleration of the electrons in them.
A high $\gamma_{\rm max}$, is, consequently, associated with particles which encountered a shock recently and have been thus accelerated. 
On the other hand, a low $\gamma_{\rm max}$ represents (macro-)particles with electrons having cooled down to lower energies. 

We find that the dynamics of the precessing jet leads to an overall broader distribution of $\gamma_{\rm max}$ as compared to 
the simulations of the steady and the variable jet. 
This can be explained by the presence of a larger number of shocks in the precessing jet \citepalias{dubey2023}, which leads to a more sustained, continuous acceleration of particles.
Effectively the repeated acceleration does not allow them to cool down efficiently to a lower $\gamma_{\rm max}$.

The existence of so many shocks in the end is responsible for producing the bump in the particle population with $\gamma_{\rm max} \simeq 10^4-10^6$ in the precessing jet, which is not present in the steady and variable jets. 
Overall, the ongoing, continuous cooling and acceleration of these {\em mid-energy} particles is responsible for the flattening of the emission spectrum for the precessing jet in the $10^{3}-10^{9}$ GHz frequency band.

On the other hand, the variable jet carries many more particles with lower maximum energy, i.e. with $\gamma_{\rm max} \simeq 10^3$. 
This  can be explained by the lower number of shocks that are present in the variable jet \citepalias{dubey2023}, leading to a suppression of particle acceleration. 
Consequently, particles once cooled down are not, in general, accelerated (and cooled) subsequently as a result of fewer shocks. 
This leads to suppression of overall emission in the variable jet.
The particles carried along with the variable jet therefore emit predominantly at lower frequencies, which results in a dominant low-frequency hump as compared to the high-energy hump, as well as a more pronounced low-frequency hump, as in comparison with the steady and the precessing jet. 
%
\section{Electron Populations}
\label{sec:population}

As discussed in Section~\ref{sec:core_spectrum}, the observed double-hump synchrotron SEDs of jets in our simulations result from the presence of distinct populations of electrons.
Observationally, these electron populations are typically distinguished by the peak frequency at which 
they radiate.

Although the double-peaked synchrotron SEDs suggest that only two of these populations dominate the core SEDs, it is also possible that the jet has more than two populations of electrons.
In fact, given the complex jet dynamics and the existence of several distinct dynamical features in the jet dynamics (i.e. various types of shocks), we actually expect the existence of various electron populations.
If true, the question arises, why only two spectral features are visible?

In this section, we will try to disentangle the different electron populations by applying two approaches.
The first approach (discussed in Section~\ref{sec:2Dpop}) is to identify the regions (and, thus, populations) which radiate at 
different peak frequencies in the 2D intensity maps in the plane of the sky. 
This method is observationally motivated, or from an observer's point of view, since observations consider integrated 2D intensity map in the plane of the sky. 
The second approach (discussed in Section~\ref{sec:3Dpop}), exploits the 3D structure of the jet that is available to us via the simulation approach.
The latter allows us, in particular, to study the SEDs from different electron populations in the 3D jet structure which we will discuss in Section~\ref{sec:epopulations}.

\subsection{Distribution of Electron Populations in 2D Plane of the Sky}
\label{sec:2Dpop}
\begin{figure*}
    \centering
    \includegraphics[width=0.32\linewidth]{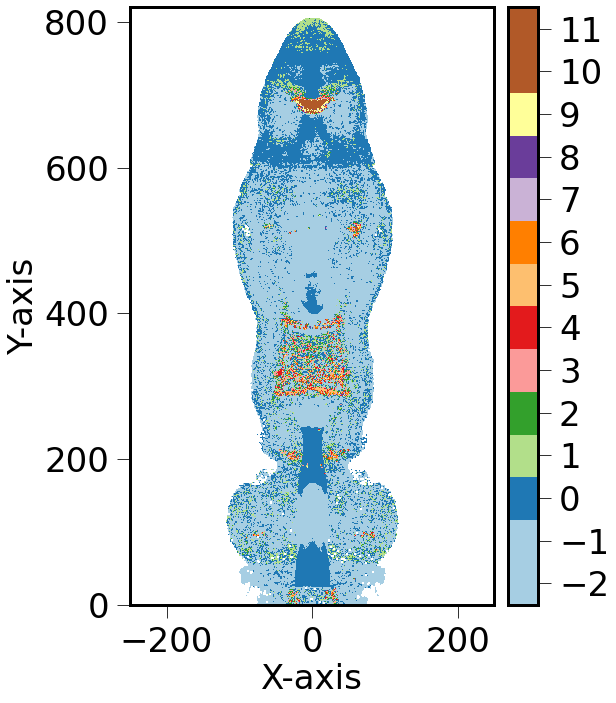}
    \includegraphics[width=0.32\linewidth]{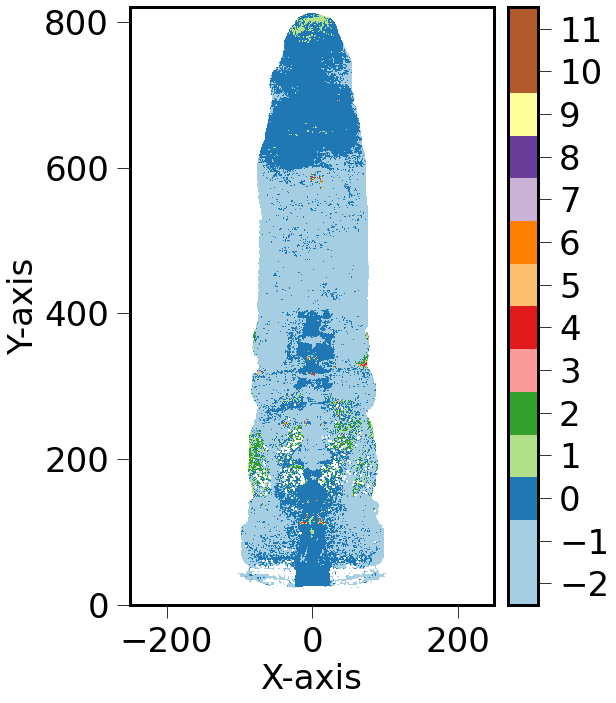}
    \includegraphics[width=0.32\linewidth]{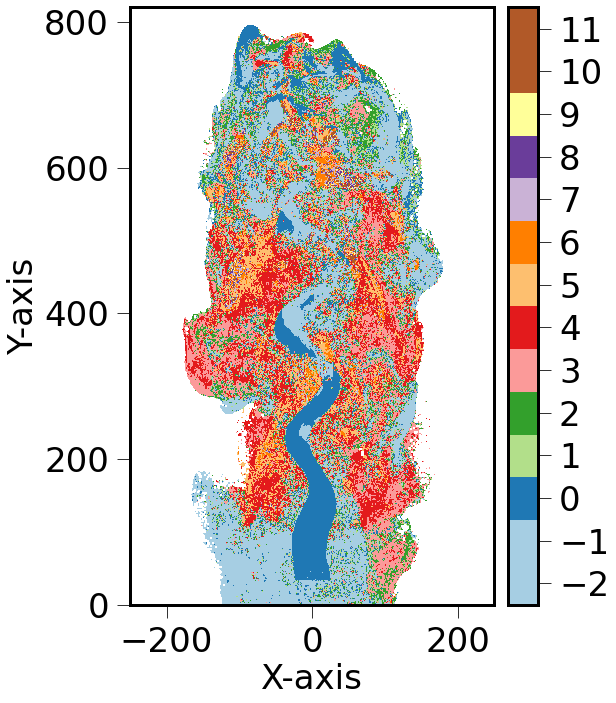}
    
    \caption{Distribution of the peak frequency at which the electrons in a grid cell radiate.
    Shown for the steady jet {\em Std90} (left), variable jet {\em Var90} (center), and precessing jet {\em Prc90} (right) at time $t=50$ 
    and for inclination to the l.o.s. $\hat{\textit{\textbf{n}}}_{\rm los} = 90^{\circ}$. 
    The colorbars shows the $\log$ of peak frequency (in GHz) of the grid cell. 
    Different peak frequencies shown here correspond to different electron populations.
    This 2D distribution is derived from the intensity map, thus the l.o.s.-integrated 3D emissivity, projected into the plane of the sky.
    }
    \label{fig:population_dist}
\end{figure*}
We now discuss the spatial distribution of the electron populations we find in our simulations in more detail.
One way of distinguishing these electron populations is the peak frequency at which they radiate.
In order to investigate this, we first integrate the 3D emissivity distribution along the l.o.s. to obtain the 2D intensity distribution in the plane of the sky (Equation~\ref{eq:intensity}).

Then, to show the locations of various electron populations in the plane of the sky, 
we find for every grid cell of the 2D intensity distribution the peak frequency for the electrons which contribute to the radiation of the (2D) grid cell. 
This {\em peak} frequency is defined as the frequency at which the radiant intensity $\nu I_{\nu}$ of that particular (2D) grid cell is the highest.

In an observational context, the peak frequency would also imply the frequency bands at which the different regions of the jet can be observed as a result of higher emission at those bands. 
Additionally, it helps in understanding which regions of the jet are responsible for the emission that is observed in different bands, such as radio, optical, or X-ray.

Since the core SEDs of different jets (Figure~\ref{fig:spc_t50}) have a similar shape for the different inclinations, it is sufficient to do this exercise for just one choice of inclination, {\em e.g.} a viewing angle of $\hat{\textit{\textbf{n}}}_{\rm los} = 90^{\circ}$.

In Figure~\ref{fig:population_dist} we show the distribution of peak frequencies at each grid cell for the
steady jet {\em Std90}, precessing jet {\em Prc90}, and variable jet {\em Var90} at time $t=50$ and for an inclination to the l.o.s. $\hat{\textit{\textbf{n}}}_{\rm los} = 90^{\circ}$. 

Overall, we find the presence of more than one population of electrons, each characterized by radiation maximum at different peak frequencies. 
In case of the steady jet, emission at the highest frequencies  $\simeq 10^{10}-10^{11}$\, GHz comes mostly from the Mach shock region with $Y \simeq 700$.
In addition, we find a few regions, although small, which are affected by the strong steady shock ($Y \simeq 270-400$) and which also emit at this frequency band.
It is the emission from these electrons that lead to the formation of the high-frequency peak in the core SEDs of the steady jet (see Figure~\ref{fig:spc_t50}).

On the other hand, the low-frequency peak located at $\simeq 1$\, GHz (shown in dark blue color), results from emission from regions {\em in} and {\em surrounding} the recollimation shock, $Y \simeq 150-250$, along with regions close to the jet base ($Y \simeq 25-60$), and also locations {\em in} and {\em surrounding} the termination shock, $Y \simeq 350-800$. 
It is interesting to note that most of the emission from the steady jet originates from these rather small regions, but dominates the core SEDs.

We find that most parts of the steady jet radiate at frequencies $\simeq 10^{-2}-10^{-1}$\, GHz (shown in light blue color).
However, the overall luminosity $\nu L_{\nu}$ at these frequencies is comparatively low, much lower than at the peak frequencies.
We find other electron populations that radiate in the $\simeq 10^{2}-10^{6}$\, GHz band as well (shown in shades of green, red, and orange colors), but their contribution to the overall SEDs also low. 
Tracing back the electrons radiating in these bands, we see that they were pre-dominantly energized by the strong steady shock.

In comparison, the intensity map of the variable jet has a more uniform structure, with the majority of the grid cells radiating  dominantly at
frequencies $\nu \simeq 10^{-2}-1$\,GHz. 
This, however, does not imply that the variable jet is not radiating at higher frequencies. 
Rather, in the variable jet the radiation from the majority of cells is enhanced at lower frequencies as compared to that for the higher frequencies. 
In particular, the high-frequency X-ray emission at $\nu \simeq 10^{9}$\,GHz arises largely from Knot A of the variable jet {\em Var90} as shown in Figure~\ref{fig:knotanalysis_VarNorm90}. 
The other knots {\em viz.} Knots B, C1, C2, D, and E also contribute to the X-ray flux of the variable jet.

This is also consistent with the SED of the variable jet (Figure~\ref{fig:spc_t50}), where we see that the high-frequency peak has a lower luminosity than the low-frequency peak as compared to the steady and the precessing jet (see our discussion in Section~\ref{sec:core_spectrum}). 
\begin{figure*}
    \centering
    \includegraphics[width=0.32\linewidth]{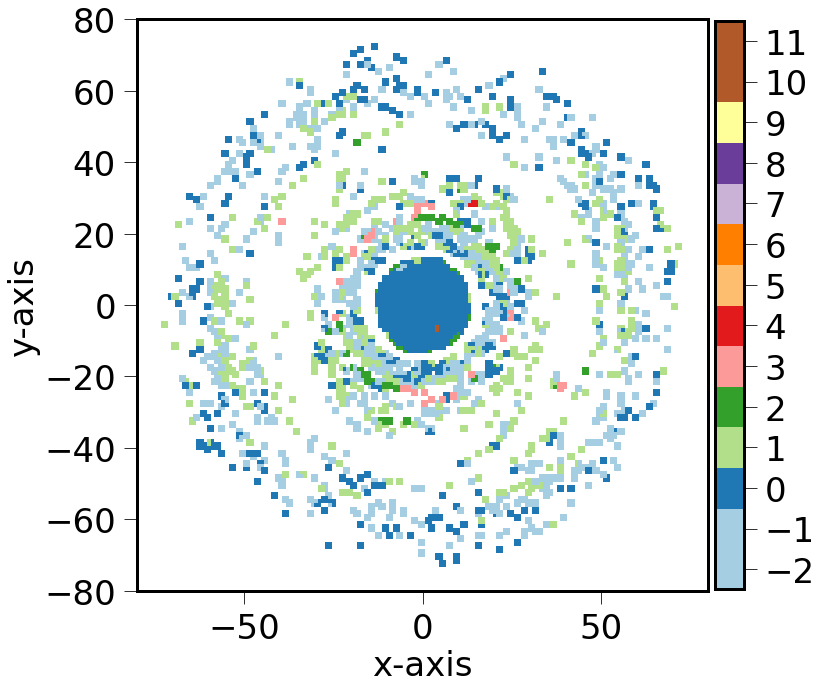}
    \includegraphics[width=0.32\linewidth]{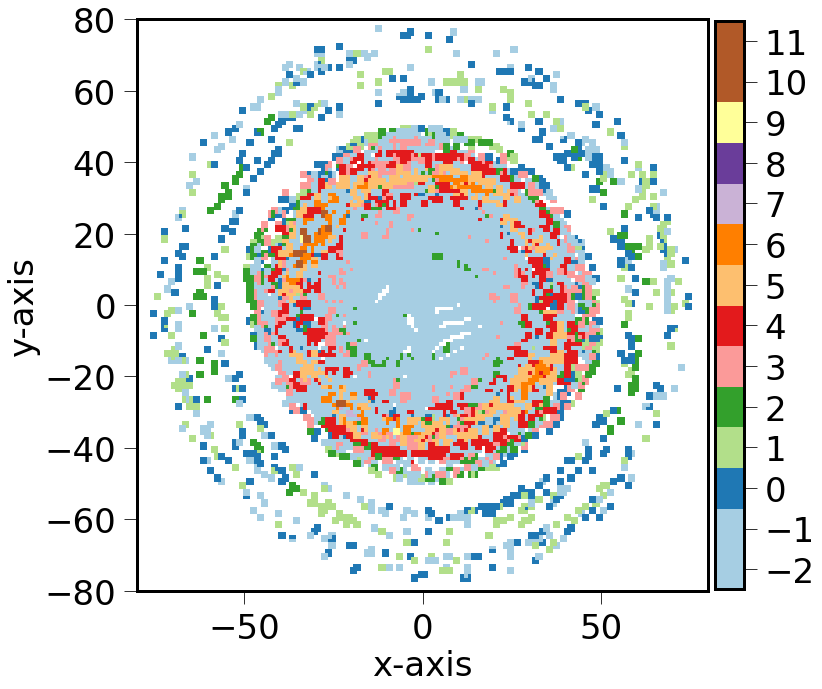}
    \includegraphics[width=0.32\linewidth]{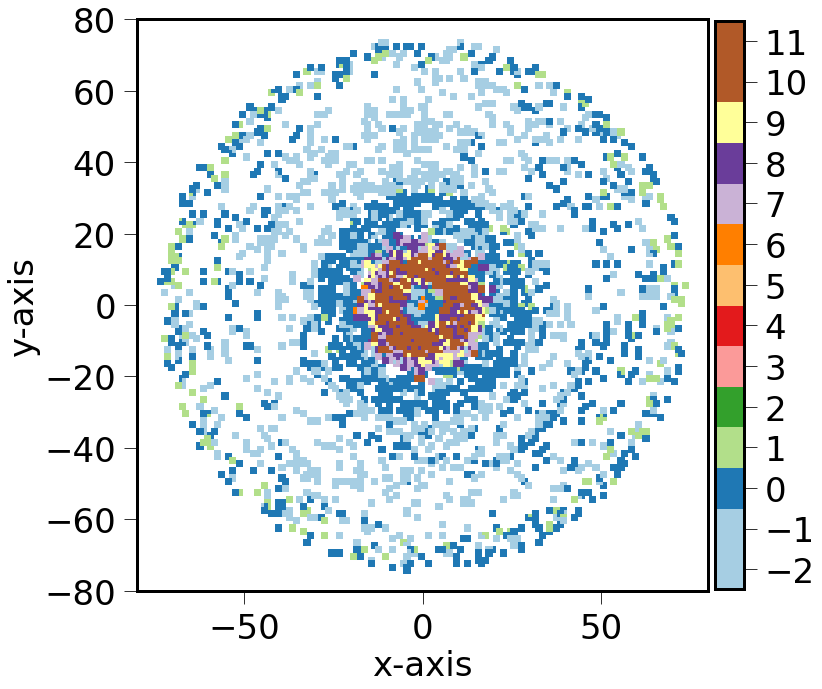}
    
    \caption{Distribution of the peak frequency at which the electrons in a grid cell radiate.
    Shown are the 2D slices of the 3D distribution at $z=220$ (recollimation shock, left), $z=390$ (strong, steady shock, center), and $z=685$ (Mach shock, right) for the steady jet {\em Std90} at time $t=50$ 
    and for inclination to the l.o.s. $\hat{\textit{\textbf{n}}}_{\rm los} = 90^{\circ}$. 
    The colorbars shows the $\log$ of peak frequency (in GHz) of the grid cell. 
    Different peak frequencies shown here correspond to different electron populations.
    These 2D slices is derived from the 3D emissivity maps ($\nu J_{\nu}$).
    }
    \label{fig:population_3Ddist}
\end{figure*}
For the precessing jet the peak frequencies are quite distributed throughout the jet structure (see second panel of Figure~\ref{fig:population_dist}).
This is naturally expected from the dynamical evolution for this jet.
The electrons are accelerated repeatedly, and frequently, as a result of the many shocks.
They further mix up with each other due to the stirring effect of dynamics caused by the precessing nozzle. 
In practice, these electrons subsequently cool down after acceleration and emit at a higher peak frequency.  
The peak at $\simeq 1$\, GHz comes from the inner jet, shown in dark blue color. 
Other electron populations emitting at $10^{2}-10^{6}$\, GHz are also present and play an important role in the core spectrum. 
It is these populations that flatten the core spectrum of the precessing jet in this band (Figure~\ref{fig:spc_t50}).

\subsection{The 3D Distribution of Electron Populations}
\label{sec:3Dpop}
Here we study the 3D distribution of electron populations.
We characterize the electron populations by the peak frequency we find for each grid cell in the 3D emissivity distribution. 
The peak frequency at a particular grid cell, in our the 3D treatment, is defined as the frequency at which the emissivity 
$\nu J_{\nu}$ is at maximum at that respective grid cell. 

In Figure~\ref{fig:population_3Ddist} we show the 2D slices of the 3D distribution of the peak frequencies across the steady jet
{\em Std90} ({\em i.e.} the
$x-y$ plane\footnote{Note that the $x-y-z$ coordinate system applied for the MHD simulations represents a fixed frame of
reference irrespective of the line of sight and is different from the $X-Y-Z$ coordinates.}) 
at different height $z$ along the jet at time $t=50$. 
We select the different heights to represent certain interesting regions in the jet {\em viz.} the recollimation shock at $z=220$, 
the strong, steady shock at $z=390$, and the Mach shock at $z=685$. 

We find that the recollimation shock radiates with highest power near frequency $\nu \simeq 10^9$ (shown by dark blue colour in 
the center region of the left panel). 
The strong steady shock, shown in the central panel, at altitude $z=390$, is known to be a site of efficient particle 
acceleration \citepalias{dubey2023}. 
As a result, we see in the radiation maps, that the electrons that have passed this shock emit at higher peak frequencies ranging 
from $\nu \simeq 10^{12-20}$. 

The radiation from the outer areas of the steady jet peaks at higher frequencies $\nu \simeq 10^{11-12}$.
This holds for all three chosen locations along the jet, thus for all three panels.
We interpret as a reason for this the enhanced shock acceleration from the backflowing jet material in the cocoon.
We want to emphasize here that the particles in the cocoon can emit at higher peak frequencies as  compared to majority of the jet, 
including the recollimation shock. 
However, as the number of particles in the cocoon is much lower than the jet, the total power radiated by these electrons is 
much lower - as seen in the SEDs (Figure~\ref{fig:spc_t50}).

\subsection{Characteristics of Different Electron Populations}
\label{sec:epopulations}
\begin{figure*}
    \centering
    \includegraphics[width=\linewidth]{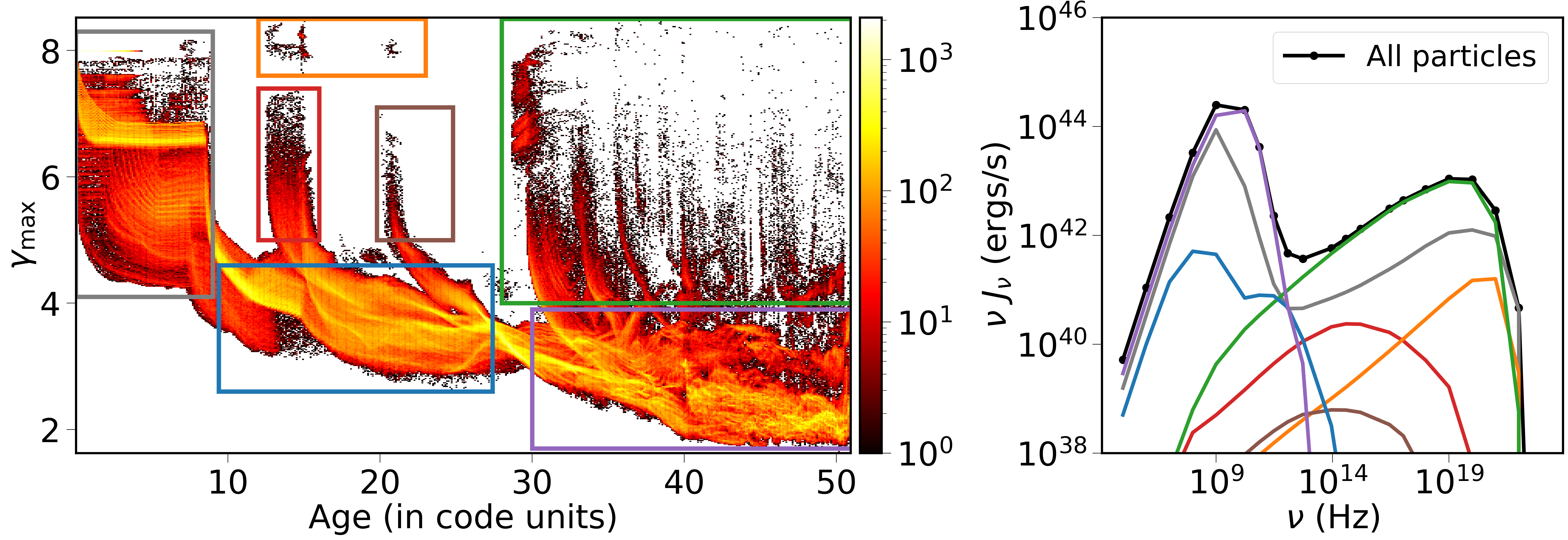}
    \includegraphics[width=\linewidth]{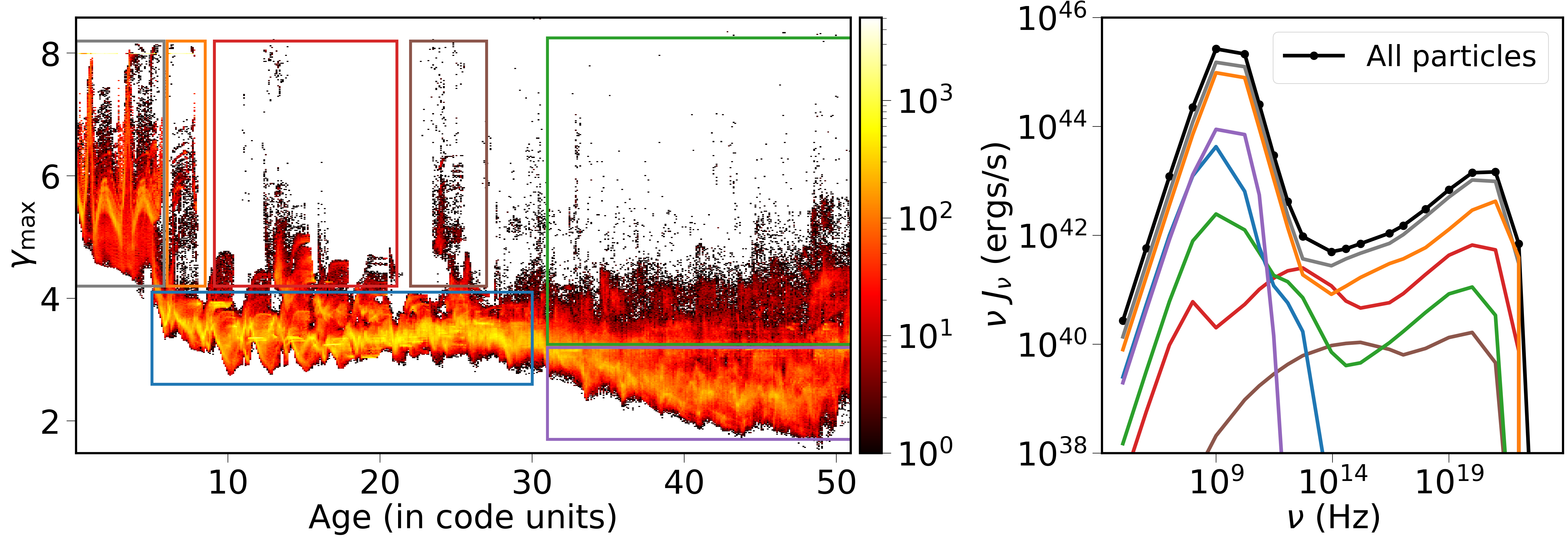}
    \includegraphics[width=\linewidth]{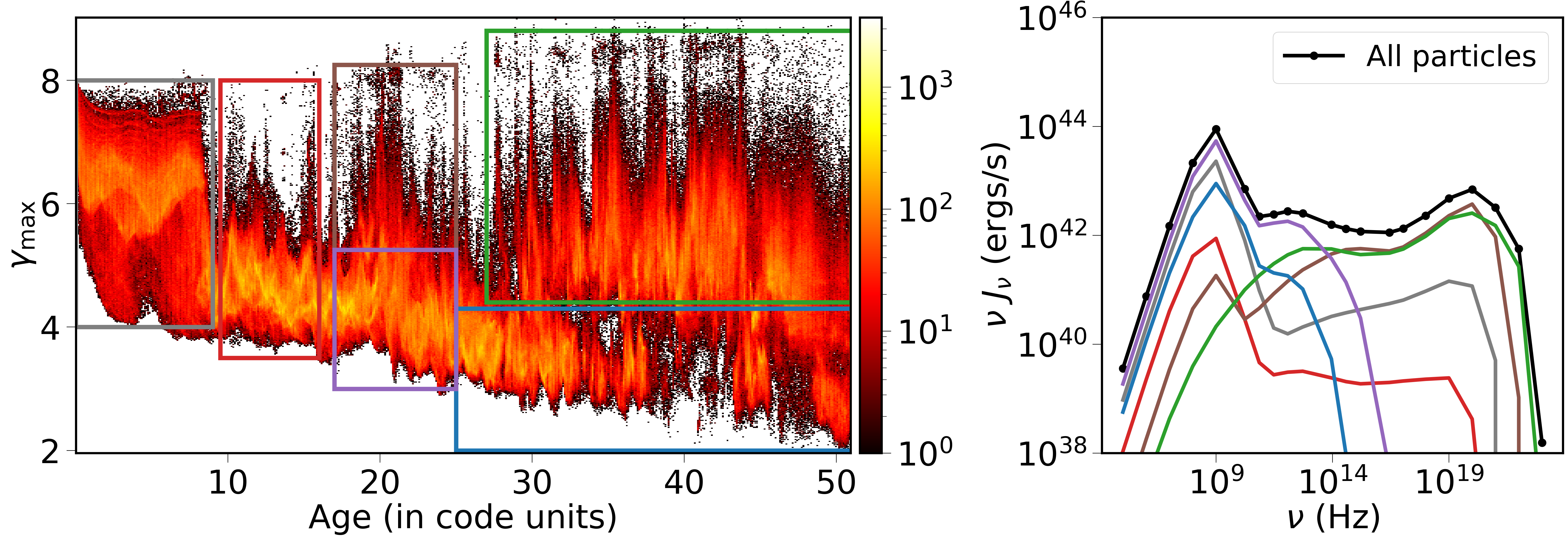}
    \caption{{\em Left:} Histogram showing $\gamma_{\rm max}$ and the age of injected particles for simulations {\em Std90} (top), {\em Var90} (center), and {\em Prc90} (bottom) at time $t=50$. The colorbars (in $\log$ scale) indicates the number of particles with a particular value of $\gamma_{\rm max}$ and 
    age. The boxes with colored boundaries therein enclose different populations of particles.
    {\em Right:} The SEDs for different populations of particles for simulations {\em Std90} (top), {\em Var90} (center), and {\em Prc90} (bottom)
    at time $t=50$. The color of the SEDs here indicates the population of those particles enclosed within the boxes of corresponding colors, 
    as shown in the left panel. 
    The black dotted line represents the {"}core{"} SED, considering all the particles in the domain.}
    \label{fig:pop_analysis}
\end{figure*}

Apart from their regional distribution and emission signatures (as discussed above), the electron populations we find may have distinct dynamical properties, as a result of their evolution, differentiating them from each other.

Hence, in order to better understand the formation of different populations of electrons, we need to disentangle the characteristics of various populations and their respective SEDs. 
For this purpose, there are two parameters of interest: 
(i) the maximum Lorentz factor $\gamma_{\rm max}$ of the electron energy distribution of an individual macro-particle, and 
(ii) the age of the macro-particle. 

We emphasize here that $\gamma_{\rm max}$ is the absolute maximum value of the Lorentz factor of electrons constituting a single macro-particle. 
Hence, it may be so that there are very few electrons (even one single electron) actually accelerated to $\gamma_{\rm max}$. 
Nevertheless, the $\gamma_{\rm max}$ of electron energy distribution does reflect the effects of present dynamical conditions on electron acceleration and radiative cooling of electrons in a macro-particle.

The first point (i)  is related to the strength of shock encountered by the macro-particle and subsequent acceleration of constituent electrons, whereas the latter point (ii) tells when in their global evolution macro-particles are accelerated by different shocks. 

In Figure~\ref{fig:pop_analysis} we present 2D histograms representing the number of (macro-)particles with their specific values of $\gamma_{\rm max}$ and their age for the different jet simulation at late evolutionary time, $t=50$. 

We see that the particles which are actually segregated in different locations (indicated by differently colored boxes) 
in fact represent unique populations\footnote{We here have identified these populations by eye, based on their $\gamma_{\rm max}$ and age.}. 
In order to connect the particle (and thus fluid) dynamics to the emission features, we have also attached the emission spectrum
of the different populations. 
We notice that the color of the curve indicates the SED, which is derived from only the particles enclosed in the box of respective color. 

For the steady jet {\em Std90}, we find that the high frequency hump in the SED at frequency $\nu \simeq 10^{10}$ GHz results predominantly from older particles (enclosed by the green box and their spectra shown by green curve) with age $\gtrsim 30$ that are accelerated by (many) shocks to $\gamma_{\rm max} \gtrsim 10^4$. 

In addition, also other populations with accelerated particles, with ages $\simeq 12-22$ and $\gamma_{\rm max} \gtrsim 10^4$ 
(enclosed in red, brown, and orange boxes), emit around this frequency band. 
However, their contribution to the high-frequency band remains small in comparison to the older particles.
Some young particles, i.e. with age $\lesssim 10$, that were recently injected (enclosed in gray box) also contribute to this band.

On the other side, the older particles with age $\gtrsim 30$ and $\gamma_{\rm max} \lesssim 10^4$ and that have therefore cooled down (enclosed in purple box)  are mostly responsible for forming the low-frequency peak at $\nu \simeq 1$ GHz.
In addition, some young particles (enclosed by the gray box) do contribute significantly to this band.
The particles of (intermediate) age $\simeq 10-28$ (enclosed in the blue box) are not significantly contributing to the emission from the steady jet.
Overall, the essential role of the older particles  for both the low-frequency as well as the high-frequency band of the SED, 
reflects and emphasizes that it is the {\em re-acceleration} of particles which is of crucial importance with respect to the jet emission.

The variable jet {\em Var90} shows a narrower distribution of $\gamma_{\rm max}$ and age as compared to the steady jet and the precessing jet. 
We emphasize that for this jet a significant contribution to the core SED originates from younger, but energetic particles, 
with an age $\gtrsim 10$ and $\gamma_{\rm max} \gtrsim 10^4$ as enclosed in gray and orange boxes.

When considering these {"}young{"} particles for the core SED of the variable jet, one may debate whether these particles are still more influenced by the initial conditions. 
However, when removing the contribution of these particles for the core SED, we find a similar double-hump SED, but now with a much lower luminosity compared to the SED that considers their contribution. 
In that case the low-energy peak would be formed by cool particles with $\gamma_{\rm max} \lesssim 10^4$, where as the high-energy peak is resulting from the energised particles with $\gamma_{\rm max} \gtrsim 10^4$. 
Overall, we may conclude that jets with variable injection are not very efficient emitters of synchrotron radiation.
This can again be understood to be resulting from a lower number of shocks being present, thus resulting in a lack of efficient {\em re-acceleration} of electrons. 
With that, the importance of a continuous particle re-acceleration along the jet is again highlighted. 

For the precessing jet {\em PrcNor90} we find a broader distribution concerning age and $\gamma_{\rm max}$.
This is indeed expected since the dynamical evolution of the jet is generating a large number of shocks constantly.
This keeps accelerating particles of all ages continuously. 
The low-frequency peak of the precessing jet is caused by a particle population of intermediate age, $\simeq 15-25$, and intermediately energy, $\gamma_{\rm max} \simeq 10^{3-5}$, (enclosed in purple box). 
However, a younger particle population of age $\lesssim 10$ (enclosed in gray box) also contributes to this band. 

Apart from these two populations, also old particles with age $\gtrsim 25$ and $\gamma_{\rm max} \lesssim 10^4$ have some, although minor, contribution to the emission at lower-frequency bands. 

On the other hand, the high-frequency emission from the precessing jet predominantly originates from energised particles with $\gamma_{\rm max} \gtrsim 10^4$ and with age $\simeq 20$ (brown box) or $\gtrsim 30$ (green box).
The flattening of the SED we observe in intermediate-frequency band with frequency $\nu \simeq 10^{12-18}$ is resulting mainly from particles of intermediate or slightly older age $\gtrsim 17$ (purple and brown boxes), and intermediate acceleration ($\gamma_{\rm max} \simeq 10^{4-6}$). 
The essential role of older particles for the SED of the precessing jet again demonstrates the importance of re-accelerating of electrons in regarding the jet emission.


\subsection{Characteristic Shock Spectra}
\label{sec:shockstructure}
\begin{figure*}
    \centering
    \includegraphics[width=\linewidth]{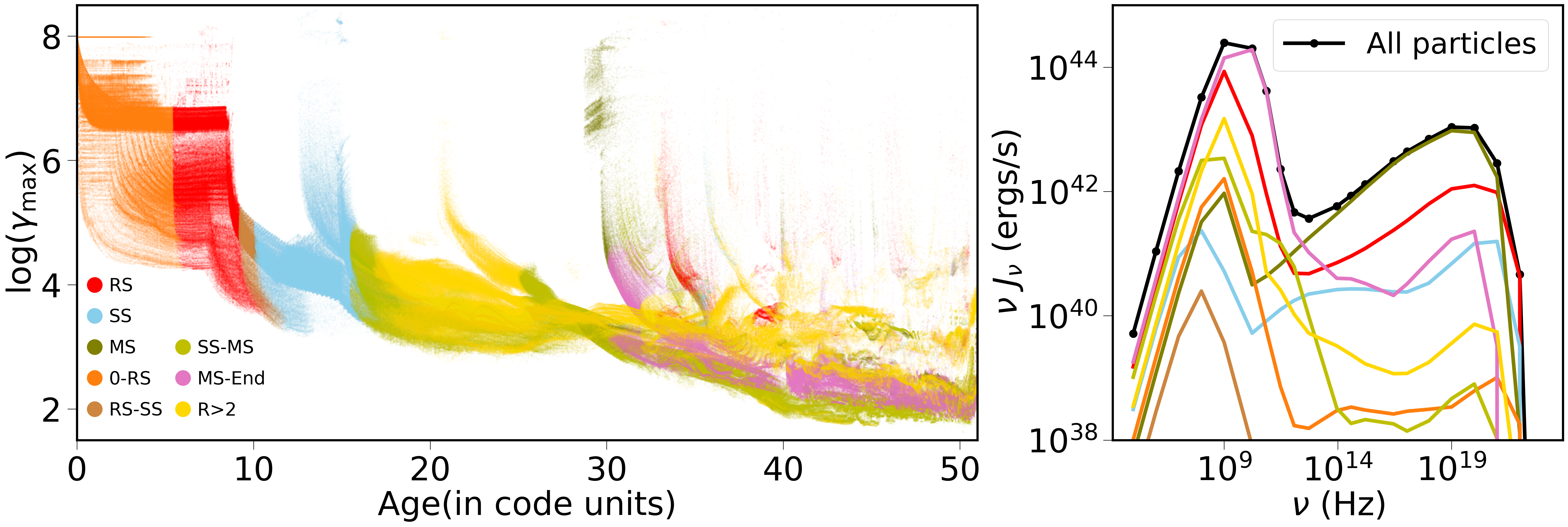}
    \caption{{\em Left:} Distribution of $\gamma_{\rm max}$ and particle age for jet simulation {\em Std90} at time $t=50$. 
    The colors indicate the location of the respective particle in the 3D jet MHD structure. 
    (i) {\em Red} indicates the recollimation shock (RS) from $z=6.4-10$, 
    (ii) {\em sky blue} indicates the strong, steady shock (SS) from $z=10.8-16$, 
    (iii) {\em dark green} indicates the Mach shock (MS) from $z=26.8-28$, 
    (iv) {\em orange} indicates the region from the base of the jet up to the recollimation shock from $z=0-6.4$, 
    (v) {\em brown} indicates the region beyond the recollimation shock up to the strong steady shock from $z=10-10.8$, 
    (vi) {\em light green} indicates the region beyond the strong steady shock up to the Mach shock from $z=16-26.8$,
    (vii) {\em pink} indicates the region beyond the Mach shock to the jet termination for $z >28$, and 
    (viii) {\em yellow} indicates the particles in the cocoon and the backflow for radii $R>2$.
    {\em Right:} The SEDs for the different particle populations, as shown on the left. 
    The color of the SEDs here indicates the same particle populations as shown left. 
    The black dotted line represents the cumulative SED, considering all the particles in the domain, thus the {"}core{"} SED.} 
    \label{fig:secpop_analysis}
\end{figure*}
Besides understanding the characteristics of different particle populations, as we discussed in last section, it is also of interest to 
understand from which regions these populations originate from, and possible what kind of different shocks produce them. 
Indeed, in our dynamical jet structure we find different kind of shock structures that are present such as recollimation shocks, the Mach
shock, the termination shock region, or the strong steady shock we.
Thus, the question arises whether these different shock structures lead to characteristically different spectral properties of the particles accelerated by them. 
This is observationally very interesting, as it would allow us to disentangle from the observed spectral properties of a jet knot 
information about the nature of the underlying shocks, which then further may provide information about the local jet fluid dynamics. 

For this purpose, here we discuss the SEDs from these shocks. 
Specifically, in Figure~\ref{fig:secpop_analysis} we show the distribution of $\gamma_{\rm max}$ and age of the particles in different 
regions of the jet along with their SEDs at time $t=50$. 
For this exercise, we choose the steady jet simulation {\em Std90}, since the various kinds of shocks are clearly and distinctly visible. 

Note that different colors in the figure correspond to different regions and shocks which can be localized in the 3D jet structure. 

We remind here that as the particles in our approach are Lagrangian (i.e. they are frozen into the underlying fluid), the time evolution 
of particles also corresponds to their spatial evolution and vice-versa. 
Therefore, as time increases, particles injected from the jet base will generally move downstream along the jet. 
Occasionally, after experiencing a shock, the particles can be re-directed in other directions as well {\em e.g.} in the cocoon, or the backflow. 

We find that after the particles are injected with a steep power law for the electron energy distribution and $\gamma_{\rm max} = 10^8$, 
they first evolve and cool down to lower $\gamma_{\rm max}$ in the region close to jet base with $z \simeq 0-6.4$. 
As a result, the emission from these particles (shown in orange) is very low as compared to the total core emission, and constrained to low-frequency regime. 

When evolving further to age $t\simeq 5$, these particles pass the recollimation shock and parts of them becomes accelerated to higher $\gamma_{\rm max}$. 
In addition to more efficient acceleration, we also see an enhanced rate of cooling in this region. 
This enhanced cooling results from collimation, and subsequent amplification of magnetic field lines as a consequence of the recollimation shock.  
The amplified magnetic field strength further leads to lower cooling times, thus leading to an enhanced rate of cooling. 
This results in an efficient cycle of repeated acceleration and cooling of electrons.
Consequently, we see highly enhanced emission, at both the low- and high-frequency bands (shown in red color), as compared to the particles upstream, that are located near the base of the jet. 

This explanation holds for the variable jet as well. 
The gray and orange curves shown in Figure~\ref{fig:pop_analysis} for the variable jet also show enhanced emission, at
both the low- and high-frequency bands. 
The corresponding young particles also pass a recollimation shock located at the base of the variable jet, enhancing both 
the acceleration through shock as well as cooling as a result of amplified magnetic field. 
Thus, we find that a double-hump feature in the spectrum is a {\em characteristics of recollimation shocks}.

Further downstream the recollimation shock in the steady jet, when these energized particles have left the recollimation 
shock region, they cool down to even lower $\gamma_{\rm max}$. 
However, now there is no subsequent energization of electrons due to absence of further shocks in this region.
Therefore, the emission from the particles in this region is drastically reduced, as evident from their spectrum (brown color). 

After evolving to an age $\simeq 10$, the particles encounter the strong, steady shock located at $z\simeq 10.8$. 
This shock is highly efficient in accelerating particles to high $\gamma_{\rm max} \simeq 10^8$. 
Further, the stronger magnetic field here lead to cooling of electron on a smaller time scale. 
This again, like the recollimation shock, induces a cycle of acceleration and cooling of electrons.
As a result, we see enhanced emission from this region (shown in sky blue color) as compared to particles in immediately upstream region outside the strong, steady shock (in brown color).  

This also explains the reduction and increase of low- and high-frequency peak luminosity, respectively, for the steady jet 
in case of a lower magnetic field $B_c$ along the jet axis (see discussion in Section~\ref{sec:core_spectrum}).  
Essentially, the less prominent recollimation shock in the steady jet leads to lesser emission at lower frequencies. 
On the other hand, a more prominent strong steady shock results in a more intense emission at higher frequencies.

Upon leaving the strong steady shock at $z \simeq 16$, the particles cool down further. 
However, some more shock structures, although weak, can be found between the strong steady shock and the Mach shock. 
As a result, we see a sharp decrease in the high-frequency emission from particles in this region (shown in light green color), accompanied by an increase in low-energy emission. 

Interestingly, upon meeting the Mach shock at $z\simeq 26.8-28$, particles are highly energized to $\gamma_{\rm max}$ reaching again values of $\simeq 10^8$. 
The emission from particles encountering the Mach shock is, hence, in the high-frequency regime (shown in dark green color), and is responsible for formation of the high-frequency peak.
Further downstream the Mach shock, these particles cool down further and emit at lower frequencies (shown in pink). 

The particles that are located in the cocoon and the backflow, thus at radii $R>2$, constitute the mid- to old-age regime of the distribution with ages $\gtrsim 18$. 
These particles contribute as well to the low-frequency peak (shown in yellow color), as a result of numerous, but weak shocks present in this region. 

Although the particles in general follow an interlinked time and spatial evolution, we find that there are always some older particles present in the recollimation shock and strong steady shock as well.
These old particles actually belong to the cocoon or the backflow, and are accelerated when passing through the recollimation or the strong steady shocks.
Consequently, the $\gamma_{\rm max}$-age distribution of older particles is a mixture of particles belonging to different shocks. 
\\
\\

Summarizing this section, by applying the concept and a thorough analysis of different {\em particle populations} that result
from shock acceleration, we find that typically different shock structures, in combination with a variation in the jet dynamics, indeed lead to unique and characteristic emission signatures. 
These findings may further augment the observational studies and help to understand the {\em underlying fluid} conditions from the {\em observed emission features}.
Essentially, we also find that the re-acceleration of energetic particles plays an important role in regard to the radiation and emission from relativistic jets. 

\section{Conclusions}
\label{sec:summary}
In this paper we have presented mock emission maps and spectra, considering particle acceleration and radiative cooling in a
time-dependent relativistic MHD jet, scaled as a pc-scale AGN jet.
We have applied the hybrid module of the PLUTO code \citep{mignone2007pluto, vaidya2018, mukherjee2021} to study the synthetic multi-wavelength synchrotron radiation signatures. 
The present approach extends on our previous paper \citepalias{dubey2023}, which studied the role of jet dynamics in particle acceleration. 

Again, the different jet dynamics is governed by three dynamically different injection nozzles, namely, a steady, a variable, and a precessing jet. 
A high resolution of 25 grid cells per jet radius is adapted, which allows us to resolve the fine structure of the jet including various shock structures. 
Langrangian macro-particles are injected with the fluid and move along with the fluid.
Each macro-particle is an ensemble of non-thermal electrons, initially following a steep power-law energy distribution. 
Overall, we inject $\sim 4$ million macro-particles with the jet. 

This initially injected electron energy spectrum is evolved along the MHD dynamics for each particle taking into account energization due to diffusive shock acceleration, as well as cooling due to synchrotron processes and IC scattering of background CMB photons. 
From the evolved electron energy spectrum, we model the synchrotron emissivity assuming a fully transparent medium, which can be integrated along a line of sight to give intensity distribution in the plane of the sky. 

Our hybrid approach allows us to study various observational aspects of relativistic jets through synthetic observation signatures. 
In particular, it allows us to investigate the interrelation between the jet dynamics, the particle acceleration, and emission features of the jet, such as the knot pattern. 
In the following we summarize the key results of our study.

(i) We see the formation of localized patterns of high intensity, observationally known as jet {\em knots}, for all jets investigated.
These knots we find in the steady and the variable jet located along the jet axis, while for the precessing jet they are dispersed across the jet, resulting from the time-dependent direction of injection. 
    
(ii) The number of knots is highest in the precessing jet. 
This is a pure consequence of the larger number of shocks present in the precessing jet.
    
(iii) As the jet evolves, new knot patterns are formed and further evolve.
We find existing knots merging with others as well, leading to a flaring intensity of the resultant knot. 
Alternatively, knots can also fade in intensity as a result of weakening of the associated shock as well as cooling of the particles. 

(iv) The positions of the knot patterns shift slightly when the jet is observed at different frequencies. 
This will lead to lag in the variability at different frequencies if light travel time would be considered along the line of sight.
    
(v) We find knot patterns that are stationary, or quasi-stationary (moving with very slow pattern speed), relativistic or non-relativistic, when viewed from a line of sight at $90^{\circ}$ from the jet axis. 
In general, the knots in the precessing jet move with higher speed (up to $0.98c$) than those in the steady jet (up to $0.76c$) and the variable jet (up to $0.9c$). 
Thus a time-dependent injection of the jet material leads to higher pattern speeds overall.
    
(vi) The relativistic pattern speed as seen at $90^{\circ}$ from the jet axis will result in superluminal knot motion when looked at the jet along a line of sight closer to the jet axis. 
Knots in the steady jet may reach only marginally superluminal speed, with an apparent pattern speed close to $c$. 
Certain knots in the variable and the precessing jet would move with projected superluminal pattern speed, up to $2c$ and $5c$, respectively.
    
(vii) Essentially, we find that the pattern speed of knots to be linked to both, the shock velocity and the speed of the underlying fluid. 
While the acceleration efficiency of electrons in the macro-particle is governed by the shock strength, the macro-particle - while cooling - is moving away from the shock - while radiating - with the speed of the underlying fluid.

(viii) The synthetic light curves show time variability as a result of knots that are flaring or fading, as well as a result of 
time-dependent jet injection in the case of the variable jet or the precessing jet. 
In particular, the variability in the radio band is less as compared to the optical and X-ray bands. 
   
(ix) The synthetic spectral energy distributions (SEDs) of all jets we investigated show a double-hump structure. 
As we show only the synchrotron emission in our synthetic spectra (cooling, however, considers IC as well), this clearly suggests the presence of {\em multiple populations of electrons}, radiating at different peak frequencies. 

(x) The steady jet and the variable jet emit prominently at $\sim 1$\,GHz as well as $\sim 10^{11}$\,GHz. 
Here, the steady jet radiates more efficiently at higher frequencies $\sim 10^{11}$\,GHz while the variable jet, radiates more prominently at lower frequencies $\sim 10 \ {\rm MHz}-1 \ {\rm GHz}$. 
This can be explained by the presence of more shocks in the steady jet, leading to more efficient acceleration and subsequent cooling of particles. 
Compared to both, the precessing jet, with the highest number of shocks in comparison, has a flatter SED, radiating efficiently in all bands from $1-10^{12}$\,GHz. 
    
(xi) As a major achievement, we disentangle different particle populations in the jets, in particular in the steady jet. 
We find these populations typically formed by particles accelerated by different shocks. 
Their subsequent cooling results in unique, characteristic spectra. 
As a result of the more complex jet dynamics, the different shocks and populations are not easy to identify in the variable jet and the precessing jet.

(xii) The population of particles that are accelerated by the recollimation shock emits predominantly 
at lower frequencies $\sim 1$\,GHz, along with relatively fainter high-energy emission.
The population with particles that are accelerated by the strong steady shock, emit over a range of frequencies from $\simeq 10^{1-6}$\,GHz. 
The high frequency emission at $10^{9-11}$\,GHz arises predominantly from the population with particles that are accelerated to very high Lorentz factor $\simeq 10^8$ by the Mach shock.
 
(xiii) There can be situations when a particle population that is newly injected can dominate the SED. 
We find this for the variable jet.
Apart from this population of younger particles, population with old particles which have cooled down contribute significantly to the low-frequency peak. 
The population with intermediately aged particles that are highly accelerated has a significant contribution to the  high-frequency peak.

(xiv) For the precessing jet we find that it is the population with intermediate and old age particles that have cooled down, 
which forms the low-frequency peak, while population with particles of intermediate and old age forms the high-frequency hump.
    
(xv) A lower magnetic field leads generally to a weaker emission in the recollimation shock and a more prominent emission in 
the strong steady shock region, as we have found for a steady-jet model. 
In these cases, the low-frequency peak luminosity reduces, while the high-frequency peak luminosity increases.

(xvi) In general we find that the population of older particles always contribute significantly to the different frequency bands. 
Essentially, this emphasizes once more the role of re-acceleration of electrons which is of extreme importance regarding the radiation from AGN jets.

In summary, we performed a detailed and thorough analysis of the dynamics and the emission signatures of pc-scale AGN jets.
We hope that our results will help in bridging the gap between the dynamical features and the emission pattern, 
as well as between the theory studies and the observations of these jets. 

In our future work, we plan to include further emission mechanisms, such as IC or synchrotron self-Compton, that are relevant 
primarily for higher frequency bands.
This would allow us to produce complete spectral signatures of jets, also on longer length- and time-scales.
We also propose to apply real telescope parameters to produce more realistic mock observations, in order to compare them directly to observed sources.

\begin{acknowledgements}
This project is financed through grant Deutsche Forschungsgemeinschaft (DFG): FOR5195 (see www.for5195.uni-wuerzburg.de).
R.D. acknowledges travel funds by the International Max Planck Research School for Astronomy \& Cosmic Physics at the University of Heidelberg (IMPRS-HD).
B.V. acknowledges funding by the Max Planck Partner Group located at Indian Institute of Technology Indore. 
We thank Andrea Mignone for providing and sustaining the PLUTO code.
We acknowledge productive conversations with Karl Mannheim, Christoph Pfrommer, Brian Reville, and Ainara Saiz P\'erez. 
All simulations were performed at the MPCDF computing center of the Max Planck Society in Garching (utilizing the compute cluster Raven).
We acknowledge a timely and very helpful referee report, also pointing out a few inconsistencies,
that improved the overall clarity of our paper.
\end{acknowledgements}

\bibliography{main}{}
\bibliographystyle{aasjournal}

\counterwithin{figure}{section}

\end{document}